\documentclass[a4paper,fleqn,9pt]{cas-dc}

\usepackage[numbers,sort&compress]{natbib}
\usepackage{doi}
\usepackage{xurl}

\usepackage{footnote}
\usepackage{xcolor}
\usepackage{makecell}
\usepackage{threeparttable}
\usepackage{tabularx}
\usepackage{siunitx}
\newcolumntype{C}{>{\centering\arraybackslash}X}
\newcolumntype{L}{>{\raggedright\let\\\tabularnewline\arraybackslash}X}
\newcolumntype{R}{>{\raggedright\let\\\tabularnewline\arraybackslash}X}

\usepackage{supertabular}   
\usepackage{subcaption}
\usepackage{graphicx}
\usepackage{hyperref}
\usepackage{float}

\usepackage{soul}
\usepackage{algorithm}
\usepackage{algpseudocode}

\usepackage{placeins}

\usepackage[intoc]{nomencl}
\usepackage{ifthen}
\usepackage{framed} 
\usepackage{multicol} 

\renewcommand{\nomgroup}[1]{%
    \item[\textbf{%
        \ifthenelse{\equal{#1}{A}}{Abbreviations}{%
        \ifthenelse{\equal{#1}{E}}{Sets and indices}{%
        \ifthenelse{\equal{#1}{P}}{Parameters}{%
        \ifthenelse{\equal{#1}{V}}{Variables}{%
        \ifthenelse{\equal{#1}{Z}}{Subscripts}{}}}}}}]%
}

\makenomenclature 

\def\tsc#1{\csdef{#1}{\textsc{\lowercase{#1}}\xspace}}
\tsc{WGM}
\tsc{QE}

\begin{document}
\let\WriteBookmarks\relax
\def\floatpagepagefraction{1}
\def\textpagefraction{.001}

\shorttitle{Optimal Coordination of Heat Pump Demand Flexibility and Battery Energy Storage Considering the Operation of Distribution Networks }

\shortauthors{Gustavo L. Aschidamini et al.}  

\title [mode = title]{Optimal Coordination of Heat Pump Demand Flexibility and Battery Energy Storage Considering the Operation of Distribution Networks}  



%

\author[1]{Gustavo L. Aschidamini} 
\cormark[1]
\ead{gla113@sfu.ca} 
\credit{Conceptualization, Data curation, Formal analysis, Investigation, Methodology, Software, Validation, Visualization, Writing – original
draft, Writing – review \& editing}

\author[1]{Guilherme S. Castiglio}
\credit{Supervision, Writing – review \& editing}

\author[1]{Mariana Resener}
\credit{Funding acquisition, Project administration, Supervision, Writing – review \& editing} 

\affiliation[1]{organization={School of Sustainable Energy Engineering, Faculty of Applied Sciences, Simon Fraser University},
            city={Surrey},
            state={BC V3T 0N1},
            country={Canada}}

\cortext[1]{Corresponding author}



\begin{abstract}
    Electrifying space heating with heat pumps (HP) and backup electric-resistance heaters (ERH) is incentivized to support decarbonization but increases voltage and thermal issues in distribution networks. Direct load control (DLC) of smart thermostats (ST) provide grid support, yet, to the best of our knowledge, existing methods do not co-optimize this flexibility with grid-scale battery energy storage systems (BESS) and voltage-regulation devices, and do not use ERH for preheating, limiting peak demand reduction and undervoltage mitigation. We propose a mixed-integer linear programming (MILP)-based model predictive control (MPC) framework that coordinates ST setpoints for preheating and setback, grid-scale BESS, on-load tap changers, switched capacitor banks, and step voltage regulators. The framework is evaluated on 23-node and 733-node distribution networks, including a real-data-based feeder, across multiple realized days. On the 23-node feeder with 50\% HP penetration and 50\% DLC participation, the proposed framework reduces the feeder peak by 23.1\%, and by 27.1\% with the co-optimized BESS. Rolling-horizon setpoint optimization contributes 6.5 percentage points beyond a day-ahead ST schedule (16.6\%), and dispatching the ERH for preheating adds 2.7 percentage points beyond backup-only operation (20.4\%). On the 733-node feeder, where thermal congestion is the binding constraint, the framework reduces overloaded branches by 94\% and overloaded transformers by 47\%. These results demonstrate the potential of coordinated demand flexibility and voltage regulation to accommodate heating electrification.
\end{abstract} 





\maketitle

\fontsize{9}{11}\selectfont


\nomenclature[E]{$T$}{Set of time steps in the prediction (look-ahead) horizon}
\nomenclature[E]{$\psi^{b}$}{Set of network branches}
\nomenclature[E]{$\psi^{n}$}{Set of network nodes}
\nomenclature[E]{$\Psi^{e}$}{Set of end-users}
\nomenclature[E]{$\mathcal{S}$}{Set of electric-resistance heater switching stages}
\nomenclature[E]{$\Omega_d$}{Set of historical outdoor-temperature profiles}

%
\nomenclature[P]{$C^{\text{kwh}}$}{Energy-loss price [\$/kWh]}
\nomenclature[P]{$C^{\text{v}}$}{Voltage-violation price [\$/pu]}
\nomenclature[P]{$\Delta t$}{Time-step length [h]}
\nomenclature[P]{$V^{\text{min}},V^{\text{max}}$}{Minimum/maximum allowed nodal voltage [pu]}

\nomenclature[P]{$R_{i}$}{Resistance of branch $i$ [pu]}
\nomenclature[P]{$X_{i}$}{Reactance of branch $i$ [pu]}
\nomenclature[P]{$f_{i}^{\text{Re,max}},f_{i}^{\text{Im,max}}$}{Maximum real and imaginary current parts in branch $i$ [pu]}
\nomenclature[P]{$a_{k},b_{k}$}{Slope and intercept of tangent line $k$} 
\nomenclature[P]{$N^{\text{LC}}$}{Number of linear (tangent-line) constraints}
\nomenclature[P]{$x_0$}{Tangent point of a piecewise-linear segment}

\nomenclature[P]{$\mathbf{S}$}{Node-branch incidence matrix}
\nomenclature[P]{$K_{i,t}$}{Adjustment factor of branch $i$ at time $t$}
\nomenclature[P]{$\mathbf{d}_{t}$}{Vector of load currents at time $t$ [pu]}
\nomenclature[P]{$\mathbf{q}_{t}^{\text{fcb}}$}{Reactive-current injection of the fixed capacitor bank at time $t$ [pu]}

\nomenclature[P]{$T_{\text{min}}^{\text{ind}},T_{\text{max}}^{\text{ind}}$}{Minimum/maximum indoor-temperature bounds [\textdegree C]}
\nomenclature[P]{$T_{t}^{\text{out}}$}{Outdoor temperature at time step $t$ [\textdegree C]}
\nomenclature[P]{$C^{\text{sh}}$}{Thermal capacitance for SH [kWh/K]}
\nomenclature[P]{$R^{\text{sh}}$}{Thermal resistance for SH [K/kW]}
\nomenclature[P]{$COP_{t}^{\text{sh}}$}{COP of the HP for SH at time $t$}
\nomenclature[P]{$COP^{\text{erh}}$}{COP of the ERH}
\nomenclature[P]{$\gamma$}{Minimum HP modulation fraction}
\nomenclature[P]{$P_{e,\text{hp}}^{\text{cap}},P_{e,\text{erh}}^{\text{cap}}$}{HP and ERH rated (maximum) power capacity of end-user $e$ [kW]}
\nomenclature[P]{$P^{\text{stage}}$}{ERH switching stage size [kW]}
\nomenclature[P]{$T^{\text{th}}$}{ ERH outdoor-temperature activation threshold [\textdegree C]}
\nomenclature[P]{$T_{e,t}^{\text{ref}}$}{Reference indoor-temperature target of non-DLC end-user $e$ at time $t$ [\textdegree C]}

\nomenclature[P]{$Q^{\text{hp}}$}{HP heat output [kW]}
\nomenclature[P]{$P^{\text{hp}}$}{HP electrical input power [kW]}
\nomenclature[P]{$\alpha_c$}{Correction factor of the COP}
\nomenclature[P]{$a_{0},a_{1},a_{2}$}{Constant, linear, and quadratic coefficients of the COP regression}
\nomenclature[P]{$\Delta T$}{Temperature difference between heat-sink and heat-source temperatures [\textdegree C]}
\nomenclature[P]{$T_{t}^{\text{sink}},T_{t}^{\text{source}}$}{Heat-sink and heat-source temperatures [\textdegree C]}  

\nomenclature[P]{$SOC_{m}^{\text{min}},SOC_{m}^{\text{max}}$}{Minimum/maximum state of charge [pu]}  
\nomenclature[P]{$SOC^{\text{init}}$}{Initial state of charge [pu]}
\nomenclature[P]{$SOC_{\text{nom}}$}{Nominal (rated) energy capacity of the BESS [pu]}
\nomenclature[P]{$\eta_{m}^{\text{BESS}}$}{Charging/discharging efficiency of the BESS at node $m$}
\nomenclature[P]{$\beta_{m}^{\text{BESS}}$}{Self-discharge rate of the BESS at node $m$}
\nomenclature[P]{$f_{m}^{\text{max,BESS}}$}{Maximum current of the BESS at node $m$ [pu]}
\nomenclature[P]{$W_{m}^{\text{max,BESS}}$}{Maximum number of BESS charge/discharge mode changes over the horizon}
\nomenclature[P]{$\theta_k$}{Angle of supporting hyperplane $k$}

\nomenclature[P]{$Y^{\text{ce}}$}{Nominal admittance of a single capacitor element [pu]}
\nomenclature[P]{$C_{m}^{\text{scb}}$}{Number of available capacitor elements at node $m$}
\nomenclature[P]{$M$}{Sufficiently large (big-$M$) constant}

\nomenclature[P]{$V^{\text{ss,min}}$}{Substation voltage at OLTC tap position 0 [pu]}
\nomenclature[P]{$\Delta V ^{\text{OLTC}}$}{Voltage change per OLTC tap step [pu]}
\nomenclature[P]{$N_{\text{taps}}^{\text{OLTC}}$}{Maximum OLTC tap index}

\nomenclature[P]{$\tau_{m}^{\text{SVR,min}},\tau_{m}^{\text{SVR,max}}$}{Minimum/maximum SVR tap position at node $m$}
\nomenclature[P]{$\Delta a_{m}^{\text{SVR}}$}{SVR voltage tap step at node $m$ [pu/tap]}
\nomenclature[P]{$V_{m}^{\text{set,SVR}}$}{SVR reference voltage setpoint at node $m$ [pu]}
\nomenclature[P]{$\beta$}{SVR regulation bandwidth (dead band) [pu]}
\nomenclature[P]{$K_{m}^{\text{SVR}}$}{Number of bits used to encode the SVR tap position at node $m$}
\nomenclature[P]{$x_{m}^{\text{SVR}}$}{Binary indicator of SVR presence at node $m$}
\nomenclature[P]{$B^{\text{max}}$}{Maximum binary tap sum of the SVR}

%
\nomenclature[V]{$c_{t}^{\text{loss}}$}{Cost of active-power losses at time step $t$ [\$]}
\nomenclature[V]{$c_{t}^{\text{volt}}$}{Penalty cost of voltage violations at time step $t$ [\$]}
\nomenclature[V]{$z_{m,t}^{\text{up}},z_{m,t}^{\text{low}}$}{Upper- and lower-bound voltage violations at node $m$ and time $t$ [pu]}
\nomenclature[V]{$V_{m,t}$}{Nodal voltage magnitude at node $m$ and time $t$ [pu]}

\nomenclature[V]{$p_{i}^{\text{loss}}$}{Active-power losses in branch $i$ [pu]}
\nomenclature[V]{$f_{i}^{\text{Re}},f_{i}^{\text{Im}}$}{Real and imaginary parts of the current flowing through branch $i$ [pu]}
\nomenclature[V]{$f_{i}^{\text{2Re}},f_{i}^{\text{2Im}}$}{Approximations of the squared real and imaginary current parts in branch $i$ [pu]}

\nomenclature[V]{$\mathbf{V}_{t}$}{Vector of nodal voltage magnitudes at time $t$ [pu]}
\nomenclature[V]{$\mathbf{f}_{t}$}{Vector of branch currents at time $t$ [pu]}
\nomenclature[V]{$\mathbf{g}_{t}$}{Vector of current injections at time $t$ [pu]}
\nomenclature[V]{$\mathbf{q}_{t}^{\text{scb}}$}{Reactive-current injection of the SCB at time $t$ [pu]}

\nomenclature[V]{$T_{e,t}^{\text{ind,set}}$}{Indoor-temperature setpoint of end-user $e$ at time $t$ [\textdegree C]}
\nomenclature[V]{$T_{t}^{\text{ind}}$}{Indoor temperature at time step $t$ [\textdegree C]}
\nomenclature[V]{$T_{e,t}^{\text{ind}}$}{Indoor temperature of end-user $e$ at time step $t$ [\textdegree C]}
\nomenclature[V]{$p_{t}^{\text{hp}}$}{HP power output for SH at time $t$ [kW]}
\nomenclature[V]{$p_{e,t}^{\text{hp}}$}{HP power output of end-user $e$ at time $t$ [kW]}
\nomenclature[V]{$p_{t}^{\text{erh}}$}{ERH power output at time $t$ [kW]}
\nomenclature[V]{$p_{e,t}^{\text{erh}}$}{ERH power output of end-user $e$ at time $t$ [kW]}
\nomenclature[V]{$p_{e,t}^{\text{hp,req}}$}{Setpoint-tracking (required) HP power of non-DLC end-user $e$ at time $t$ [kW]}
\nomenclature[V]{$p_{e,t}^{\text{def}}$}{Residual thermal-deficit power of non-DLC end-user $e$ at time $t$ [kW]}
\nomenclature[V]{$u_{e,t}^{\text{hp}}$}{Binary on/off state of the HP of end-user $e$ at time $t$}
\nomenclature[V]{$u_{e,s,t}^{\text{erh}}$}{Binary state of stage $s$ of the ERH of end-user $e$ at time $t$}

\nomenclature[V]{$SOC_{m,t}$}{State of charge at node $m$ and time $t$ [pu]}
\nomenclature[V]{$f_{m,t}^{\text{ch,Re,BESS}},f_{m,t}^{\text{dc,Re,BESS}}$}{Real charging and discharging currents of the BESS at node $m$ and time $t$ [pu]}
\nomenclature[V]{$f_{m,t}^{\text{Re,BESS}}$}{Net real current of the BESS at node $m$ and time $t$ [pu]}
\nomenclature[V]{$f_{m,t}^{\text{Im,BESS}}$}{Imaginary current of the BESS at node $m$ and time $t$ [pu]}
\nomenclature[V]{$w_{m,t}$}{Binary charging/discharging mode of the BESS at node $m$ and time $t$}
\nomenclature[V]{$\sigma_{m,t}^{\text{BESS}}$}{BESS mode switching indicator}

\nomenclature[V]{$q_{m,t}^{\text{scb}}$}{Reactive-current injection of the SCB at node $m$ and time $t$ [pu]}
\nomenclature[V]{$c_{m,l,t}$}{Binary variable for SCB activation at node $m$, level $l$, and time $t$}

\nomenclature[V]{$V^{\text{ss}}_t$}{Substation voltage magnitude at time $t$ [pu]}
\nomenclature[V]{$\tau_{t}^{\text{OLTC}}$}{Integer OLTC tap position at time $t$}
\nomenclature[V]{$\delta_{k,t}$}{Binary variable for OLTC tap $k$ selection at time $t$}

\nomenclature[V]{$\tau_{m,t}^{\text{SVR}}$}{Integer SVR tap position at node $m$ and time $t$}
\nomenclature[V]{$a_{m,t}^{\text{SVR}}$}{SVR voltage ratio at node $m$ and time $t$ [pu]}
\nomenclature[V]{$V_{n,t}$}{Voltage at the SVR regulation node $n$ at time $t$ [pu]}
\nomenclature[V]{$V_{k,t}$}{Upstream voltage at time $t$ [pu]}
\nomenclature[V]{$f_{mn,t}^{\text{Re}},f_{mn,t}^{\text{Im}}$}{Real and imaginary current parts in downstream branch $mn$ at time $t$ [pu]}
\nomenclature[V]{$\Delta V_{m,t}$}{SVR voltage contribution at node $m$ and time $t$ [pu]}
\nomenclature[V]{$\Delta f_{t}^{\text{Re,SVR}},\Delta f_{t}^{\text{Im,SVR}}$}{SVR real- and imaginary-current compensation at time $t$ [pu]}
\nomenclature[V]{$r_{h,m,t}$}{Binary variable of bit $h$ of the SVR tap encoding at node $m$ and time $t$}
\nomenclature[V]{$j_{m,t}^{\text{up}},j_{m,t}^{\text{down}}$}{Binary upper- and lower-saturation indicators of the SVR at node $m$ and time $t$}

\nomenclature[V]{$T^{\text{dev}}$}{Temperature deviation index [\textdegree C]}
\nomenclature[V]{$P^{\text{dev}}$}{Deviation period index [\%]}
\nomenclature[V]{$C^{\text{dev}}_{t}$}{Binary indoor-temperature deviation indicator at time $t$}

\nomenclature[Z]{$t$}{Time step}
\nomenclature[Z]{$d$}{Day}
\nomenclature[Z]{$m$}{Node}
\nomenclature[Z]{$n$}{Downstream node}
\nomenclature[Z]{$i$}{Branch}
\nomenclature[Z]{$e$}{End-user}
\nomenclature[Z]{$s$}{Electric-resistance heater switching stage}
\nomenclature[Z]{$l$}{Capacitor-bank operating level}
\nomenclature[Z]{$k$}{Tangent-line / tap / hyperplane index}
\nomenclature[Z]{$h$}{Bit index of the binary tap encoding}

\printnomenclature

\section{Introduction}
\label{sec:introduction}

Electrifying building heating with heat pumps (HP) in cold climates is pushing distribution networks toward their operating limits and motivating distribution system operators (DSO) to reward customers for granting direct load control (DLC) of their heating systems through internet-enabled smart thermostats (ST). Distribution utility programs in Canada, such as BC Hydro's Peak Saver in British Columbia~\cite{bchydro_peak_saver} and Independent Electricity System Operator's Save on Energy Peak Perks program \cite{peak_perks_ontario} adjust heating-system ST to reduce demand during peak events, and Yukon Energy Peak Smart Pilot~\cite{yukon_energy_program} shifts winter morning and evening peak demand from ST. These programs demonstrate that controlling end-user ST reduce heating demand during peak periods while providing financial incentives to customers. 

The value of heating flexibility to the DSO lies not only in reducing peak demand but in coordinating it with the voltage-regulation assets and grid-scale battery energy storage systems (BESS) the DSO operates. Widespread heating electrification can cause undervoltage and line congestion in distribution networks~\cite{Damianakis2025}, which the DSO must mitigate during operation to defer costly grid reinforcement~\cite{Mascherbauer2025}. A study for British Columbia, Canada~\cite{Knittel2024}, finds that heating electrification markedly raises peak demand and load variability, motivating system flexibility in cold climates. However, controlling heating demand alone does not fully exploit the flexibility available to the DSO, which also operates voltage-regulation assets, including substation on-load tap changer (OLTC), capacitor banks (CB), step voltage regulator (SVR), and can increasingly access grid-scale BESS~\cite{CER2025EnergyStorage}. Prior work has approached this control problem along four progressively broader lines: (i) aggregating HP or heating, ventilation and air-conditioning (HVAC) flexibility in isolation from the distribution network representation; (ii) embedding the network so that voltage shapes the load flexibility; (iii) adding grid-scale BESS to that flexibility; and (iv) coordinating the utility voltage regulation assets with demand response (DR), the least explored and the focus of this work. 

The first line of methods schedules aggregated HP or HVAC loads for DR by exploiting the buildings' thermal inertia as virtual energy storage~\cite{Xu2026128047}, but represents the loads in isolation from the representation of the distribution network. These works pursue demand-side objectives through varied formulations: minimizing heating costs under time-of-use (TOU) pricing while suppressing rebound~\cite{Liang2024}, maximizing aggregator revenue or minimizing a load-serving entity's reward while peak-shaving and preserving thermal comfort~\cite{Hu2018497,Wei2024}, tracking a virtual power plant (VPP) DR signal across aggregated HP~\cite{Wang20232289}, and minimizing district peak demand by coordinating HP with thermal energy storage~\cite{Vivian2020}. These methods span mixed-integer linear~\cite{Vivian2020} and mixed-integer quadratic~\cite{Hu2018497} programs to distributed  control~\cite{Wang20232289}, receding-horizon model predictive control (MPC)~\cite{Liang2024}, and metaheuristic optimization~\cite{Wei2024}; yet, none considers the distribution network constraints. 

The second line of methods embeds the network, dispatching flexible loads through network-constrained optimization such as optimal power flow (OPF), local rules, or reactive power control so that voltage limits included into the approach~\cite{Aschidamini2025review}. Centralized and community level OPF formulations coordinate HP and other flexible resources against network limits: a nonlinear OPF-based MPC schedules electric vehicles (EV), HP, thermal storage, and photovoltaics (PV) with load shedding to counter peak rebound while preventing voltage drops and branch overloading~\cite{Srithapon2023}, and a centralized quadratic OPF in a linearized branch-flow model enhances HP hosting capacity by minimizing losses~\cite{Zhan2022364}. Other methods act locally or through specific actuators. An autonomous rule-based controller enforces nodal voltage limits by controlling EV charging and HP operation in a low-voltage (LV) network~\cite{Sinha2020}, whereas a distributed MPC regulates voltage through the thermal inertia and reactive-power compensation of HVAC systems~\cite{Dhulipala20197041}. A mixed-integer linear programming (MILP)~\cite{Ellemund2026} evaluates interruption- and temperature-based DLC contracts for HP to reduce load shedding and congestion. Across these formulations, spanning nonlinear and mixed-integer programs, rule-based control, and receding-horizon MPC, voltage is regulated through load-side flexibility alone, leaving the utility's own voltage regulation outside the optimization. 

The third line of methods incorporates grid-scale BESS within the demand-side flexibility, coordinating controllable thermal loads with storage against network limits. The grid-scale BESS appear as shared community storage paired with HVAC-cooling flexibility in a day-ahead optimization using conic programming~\cite{Tascikaraoglu2019137}, within a centralized rule-based scheme spanning HP, PV, and EV~\cite{Sengor2023}, or as a VPP resource coordinated directly with the DSO without demand-side thermal loads~\cite{Liu2024}. In none of these works, the BESS operation is co-optimized with the utility's voltage regulation devices and the ST setpoint flexibility of HP. 

The fourth line of methods coordinates the utility's voltage regulation assets with DR, and it remains the least developed. The two closest works to the present work jointly optimize an OLTC and switched CB (SCB) with demand-side flexibility, as one couples distributed energy resource (DER) dispatch and a TOU program in a risk-averse optimization formulation~\cite{Gholami2023}, while the other work shifts home-appliance loads alongside EV storage and renewables in a distributed, real-time game-theoretic scheme~\cite{Tushar20184077}. In both, the demand side is treated as generic price- or load-shifting flexibility, not the thermal-setpoint control of HP households and comfort limits, while SVR and grid-scale BESS are not considered. 

Controlling ST for space heating (SH) and grid-scale BESS operation is a time-coupled problem that motivates a rolling horizon MPC approach that accounts for forecasts and handles uncertainty. The building thermal mass stores heat across intervals and grid-scale BESS state of charge links to charging and discharging over the day, so preheating and setback of the ST and the BESS operation are valuable when planned against forecasted feeder loading and weather conditions. Rule-based and real-time schemes~\cite{Sinha2020,Sengor2023,Tushar20184077} respond to the current state and cannot anticipate peak demand or optimize solution to preheat indoor spaces or pre-charge storage, whereas day-ahead optimization schedules~\cite{Gholami2023,Tascikaraoglu2019137,Zhan2022364} commit to a forecast-based plan with no mechanism to correct deviations during operation. Rolling horizon MPC~\cite{Dhulipala20197041,Srithapon2023} combines forecast-based planning with feedback re-optimization, enabling a formulation to adjust the control according to realized power demand and weather. 

To the best of our knowledge, no rolling-horizon MPC co-optimizes the DLC of ST setpoints for households with HP and backup electric-resistance heaters (ERH) jointly with the voltage regulation devices and grid-scale BESS. Existing methods schedule the voltage-regulation devices (OLTC and CBs) day-ahead~\cite{Resener20231932} and separately from the ST setpoint flexibility, so indoor-space preheating and temperature setback are not resolved together with voltage regulation~\cite{Zhan2022364}. The grid-scale BESS is likewise dispatched apart from the heating flexibility and the voltage-regulation devices, even though it adds energy flexibility~\cite{Locher2026144} and is being expanded, with new installations across Canada expected by 2030~\cite{CER2025EnergyStorage}. Where a backup ERH is modeled, it appears as a fixed auxiliary device~\cite{Srithapon2023} that does not participate in preheating, while its additional heating capacity can be exploited to shift power demand away from expected peak demand periods at the distribution network feeder when the HP alone cannot provide sufficient preheating. Consequently, the flexibility of the building thermal mass and of grid-scale storage is not coordinated with the distribution utility's existing assets to accommodate heating electrification.  

This paper presents an MILP-based MPC that co-optimizes thermal energy stored in buildings, activated through the ST setpoints of household HP and backup ERH for SH, together with grid-scale BESS and voltage regulation devices. Table~\ref{tab:comparison} positions the proposed method against the discussed literature, including those methods that incorporate the distribution network model. The main contributions of this work are:
\begin{itemize}
    
    \item A centralized control framework that dispatches the energy flexibility of end-user building thermal mass activated through ST setpoints, jointly with the grid's voltage regulation assets and DER operation. 
        
    \item An MILP-based MPC model that co-optimizes the DLC of ST indoor-temperature setpoints, realized as preheating and temperature setback of HP/ERH households, with the voltage regulation devices: the substation OLTC, SCB, and SVR operating under autonomous line-drop compensation (LDC). By re-solving the MILP over a forecast look-ahead horizon on the realized base-load and outdoor-temperature states, the coordination aligns end-user energy flexibility with network voltage regulation at every control step. The MILP formulation enables the computation of globally optimal solutions for the resulting coordination problem. Unlike deterministic day-ahead scheduling, which assumes perfect load and weather forecasts, this rolling-horizon update accounts for base-load and weather uncertainty without depending on robust or stochastic optimization. 

    \item A coordinated ST setpoint control formulation that enables ERH to participate in demand flexibility by activating them for additional preheating, rather than restricting their operation to backup heating for thermal comfort. The ST setpoints determine the coordinated operation of HP and ERH, allowing the ERH capacity to increase preheating flexibility and shift heating demand away from network peak periods when the HP alone cannot provide sufficient preheating.   

    \item The co-optimization of a grid-scale BESS as a flexibility resource coordinated with the demand-side thermal flexibility and the voltage regulation devices, bounded by an intra-day charge/discharge switching limit that constrains battery cycling for degradation.   
\end{itemize}

\begin{table*}[t]
\centering
\scriptsize
\setlength{\tabcolsep}{3pt}
\caption{Positioning of the proposed method against related coordination studies. \checkmark: addressed; $\sim$: partial/indirect; --: not addressed.}
\label{tab:comparison}
\begin{threeparttable}
\begin{tabular}{llcccccccc}
\toprule
\multirow{2}{*}{Reference} &
\multirow{2}{*}{\makecell{Flexible\\load}} &
\multirow{2}{*}{\makecell{Utility\\BESS}} &
\multirow{2}{*}{\makecell{Backup\\ERH}} &
\multicolumn{3}{c}{Voltage regulation devices} &
\multirow{2}{*}{Method} &
\multirow{2}{*}{\makecell{Control\\scheme}} &
\multirow{2}{*}{Network} \\
\cmidrule(lr){5-7}
 & & & & OLTC & CB/SCB & SVR & & & \\
\midrule
\multicolumn{10}{l}{\textit{Load control embedding the network}}\\
\cite{Zhan2022364}          & HP       & --            & --             & -- & -- & -- & QP  & DA       & Real, LV \\
\cite{Sinha2020}            & HP, EV   & --            & --             & -- & -- & -- & RBC  & Reactive & Real, LV \\
\cite{Dhulipala20197041}    & HVAC     & --            & --             & -- & -- & -- & QP-MPC  & RH       & Test, MV \\
\cite{Srithapon2023}        & HP, EV   & --            & $\sim$\tnote{a}& -- & -- & -- & NLP-MPC  & RH       & Test, LV \\
\cite{Ellemund2026}         & HP       & --            & --             & -- & -- & -- & MILP  & DA       & Synthetic, MV/LV \\
\midrule
\multicolumn{10}{l}{\textit{Storage coordinated with demand flexibility}}\\
\cite{Tascikaraoglu2019137} & HVAC     & $\sim$\tnote{b}& --            & -- & -- & -- & SOCP & DA       & Test, MV \\
\cite{Sengor2023}           & HP, EV   & $\sim$\tnote{c}& --            & -- & -- & -- & RBC   & Reactive & Real, MV/LV \\
\cite{Liu2024}              & DER & $\sim$\tnote{d}& --            & -- & -- & -- & MILP & DA       & Test, MV \\
\midrule
\multicolumn{10}{l}{\textit{Voltage regulation assets coordinated with DR}}\\
\cite{Gholami2023}          & TOU-DR   & $\sim$\tnote{e}& --            & \checkmark & \checkmark & -- & MILP & DA & Test, MV \\
\cite{Tushar20184077} & Appliances, EV & -- & -- & \checkmark & \checkmark & -- & Game-theory & Reactive & Real, MV/LV \\
\midrule
\textbf{This work} & \textbf{HP + ERH} & \textbf{\checkmark} & \textbf{\checkmark} &
\textbf{\checkmark} & \textbf{\checkmark} & \textbf{\checkmark}\tnote{f} & \textbf{MILP-MPC} & \textbf{RH} & \textbf{Test \& Real, MV}\\
\bottomrule
\end{tabular}
\begin{tablenotes}[flush]
\footnotesize
\item CB/SCB: (switched) capacitor bank; SVR: step voltage regulator.
Method: NLP, non-linear programming; QP, quadratic programming; RBC, rule-based controller; SOCP, second-order cone programming; OPF, optimal power flow; MPC, model predictive control; MILP, mixed-integer linear programming.
Control scheme: DA, day-ahead; RH, rolling horizon; Reactive, rule-based / no look-ahead.
Network: test / synthetic / real, and voltage level (LV, low voltage; MV, medium voltage).
\item[a] Fixed anti-freeze auxiliary heater, not staged backup.
\item[b] Shared community storage, not utility-scale.
\item[c] Battery dispatched by economic/emission signals, not co-optimized.
\item[d] Storage aggregated at the VPP/aggregator level.
\item[e] Generic energy storage with a TOU-DR program.
\item[f] SVR operated autonomously via line-drop compensation (setpoint optimized day-ahead).
\end{tablenotes}
\end{threeparttable}
\end{table*}

The framework is evaluated on a 23-node test system~\cite{Ferraz2021} and a real 733-node distribution network~\cite{Resener20231932}, with building thermal dynamics capturing the SH demand of electrified households. Benchmarking the coordinated scheme against a no-DLC voltage regulation baseline, day-ahead schedules of the ST setpoints and of the voltage regulation devices, autonomous local voltage regulation control, a backup-only ERH operation, and the addition of a grid-scale BESS isolates the value of each design choice for energy flexibility and voltage regulation. The results show that coordinating the aggregate building thermal mass with the network's voltage regulation assets and storage over a rolling horizon has the potential to lower the coincident peak demand of distribution feeders while keeping indoor temperatures within the comfort band, relieving undervoltage and thereby increasing the heat-electrification hosting capacity of the existing network.  

The remainder of this paper is organized as follows. \hyperref[sec:method]{Section~\ref*{sec:method}} introduces the proposed optimization model. \hyperref[sec:case_study]{Section~\ref*{sec:case_study}} describes the case study, including the distribution networks under study, the characteristics of the heating devices, and the simulation setup. \hyperref[sec:results]{Section~\ref*{sec:results}} presents and discusses the simulation results. Finally, \hyperref[sec:conclusions]{Section~\ref*{sec:conclusions}} concludes the paper and outlines directions for future work.

\section{Proposed Optimization Model}
\label{sec:method}

\begin{figure*}[pos=t]
    \centering
    \includegraphics[width=0.67\linewidth]{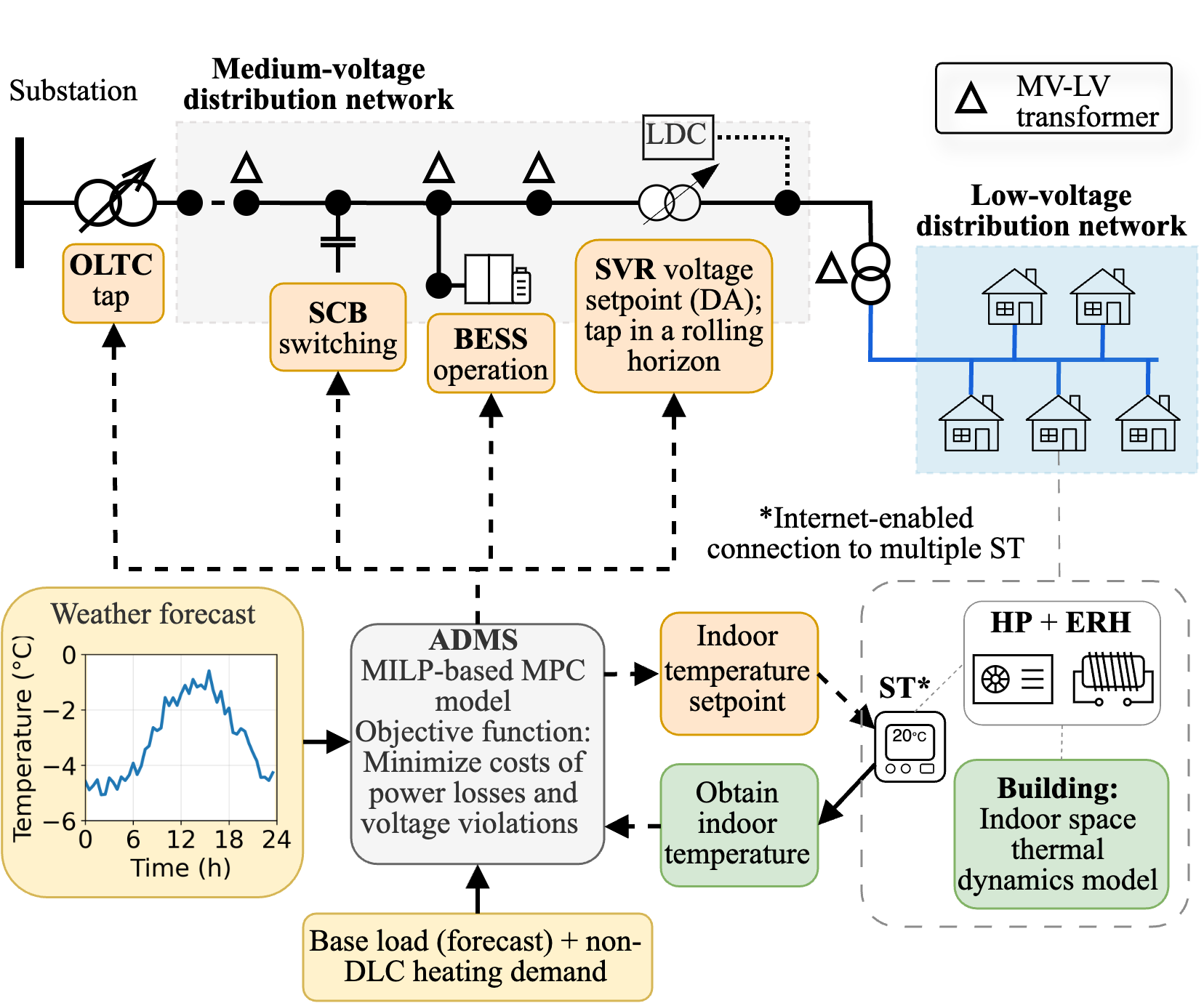}
    \caption{Overall framework of the proposed coordination of end-user ST setpoints, grid-scale BESS, and voltage-regulation devices.}
    \label{fig:overall_framework}
\end{figure*}

Fig.~\ref{fig:overall_framework} presents the proposed framework, in which an MILP-based MPC embedded in the DSO's advanced distribution management system (ADMS) coordinates end-user flexibility from ST-controlled HP and ERH with a grid-scale BESS and MV voltage regulation devices. The framework uses day-ahead forecasts of outdoor temperature and inflexible base load, together with the heating demand of non-DLC households. At each time step, the MPC sends indoor-temperature setpoints to the ST, dispatches BESS charging and discharging, and controls the OLTC tap position and SCB switching, while the SVR operates autonomously under LDC. The indoor-temperature and heating-load trajectories are obtained from a building thermal dynamics simulation. An autoregressive (AR(1)) weather model~\cite{aschidamini2026122015} generates the realized outdoor temperature used to advance the simulation, with deviations from the day-ahead forecast driving the rolling-horizon updates. 

\begin{figure}[pos=t]
    \centering
    \includegraphics[width=0.70\linewidth]{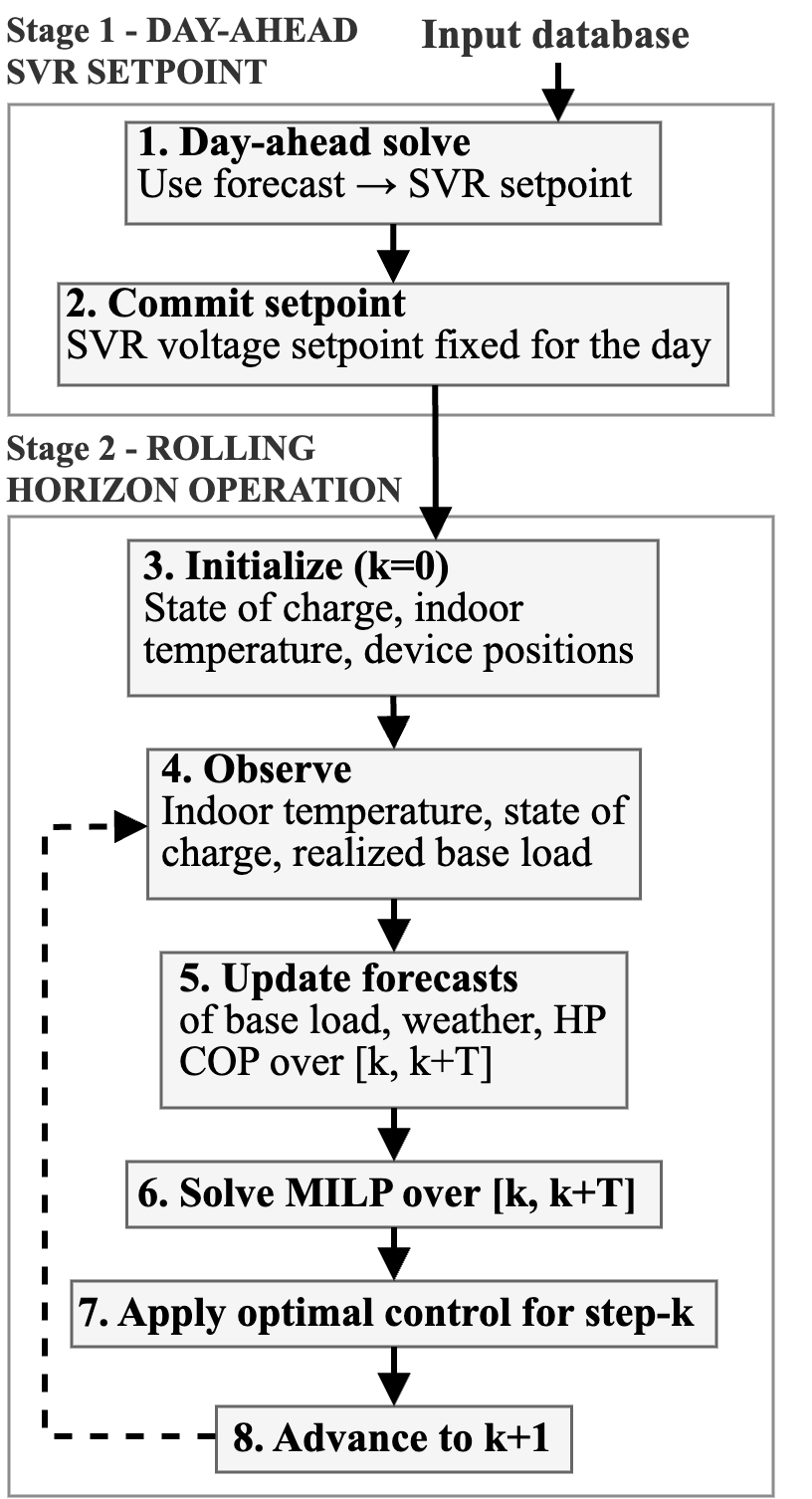}
    \caption{Stages of the proposed framework, including day-ahead SVR setpoint determination and rolling-horizon MPC operation.}
    \label{fig:flowchart_mpc}
\end{figure}

The proposed framework comprises two stages, as shown in Fig.~\ref{fig:flowchart_mpc}. In Stage~1, the SVR voltage setpoint is determined day-ahead based on the forecast and fixed for the operating day. In Stage~2, the model is re-solved every $\Delta t$ (e.g., 30~min) over a prediction horizon $T$. At each time step $k$, the realized indoor temperatures, BESS state of charge, device positions, and base load are observed, and the base-load, weather, and HP coefficient of performance (COP) forecasts are updated over $[k,\,k{+}T]$. The resulting MILP is then solved, only the first control action is implemented, and the horizon advances to $k{+}1$. This process is repeated until the end of the operating day.  

The objective function minimizes the total cost over the time horizon $T$, as defined in~\eqref{eq:obj_model}. It comprises the cost of power losses ($c_{t}^{\text{loss}}$) and the penalty cost of voltage magnitude violations ($c_{t}^{\text{volt}}$), and is adapted from \cite{Ferraz2021,Resener20231932}. The objective~\eqref{eq:obj} sums these two terms over the horizon. The loss cost~\eqref{eq:obj_loss} depends on the energy loss price $C^{\text{kwh}}$ ($\$/\text{kWh}$), the step length $\Delta t$, and the branch losses $p_{i,t}^{\text{loss}}$ over the set of branches $\psi^{b}$. The voltage-violation cost~\eqref{eq:obj_volt} penalizes, at the price $C^{\text{v}}$ (\$/pu), the violations of the upper and lower voltage bounds at each node of the set $\psi^{n}$. These violations, $z_{m,t}^{\text{up}}$ and $z_{m,t}^{\text{low}}$, are defined by~\eqref{eq:volt_upper}--\eqref{eq:volt_non_neg}~\cite{Resener2016895} as the amounts by which the nodal voltage magnitude $V_{m,t}$ exceeds the maximum voltage $V^{\text{max}}$ or falls below the minimum voltage $V^{\text{min}}$.
\begin{subequations}
\label{eq:obj_model}
\begin{align}
    \min \;\; & \sum_{t \in T} \left[\, c_{t}^{\text{loss}}
        + c_{t}^{\text{volt}} \,\right]
        \label{eq:obj} \\
    & c_{t}^{\text{loss}} = C^{\text{kwh}} \Delta t
        \sum_{i\in \psi^{b}} p_{i,t}^{\text{loss}}
        \label{eq:obj_loss} \\
    & c_{t}^{\text{volt}} = C^{\text{v}} \Delta t
        \sum_{m\in \psi^{n}} \left( z_{m,t}^{\text{up}} + z_{m,t}^{\text{low}} \right)
        \label{eq:obj_volt} \\
    & z_{m,t}^{\text{up}} \ge V_{m,t} - V^{\text{max}}
        \label{eq:volt_upper} \\
    & z_{m,t}^{\text{low}} \ge V^{\text{min}} - V_{m,t}
        \label{eq:volt_lower} \\
    & z_{m,t}^{\text{up}}, \; z_{m,t}^{\text{low}} \ge 0
        \label{eq:volt_non_neg}
\end{align}
\end{subequations}

This model adopts the linear power flow formulation from
\cite{Resener2016895} and further validated in \cite{Resener2019378,Ferraz2021}, with loads modeled as constant currents, balanced three-phase loads, and power losses approximated by piecewise linear expressions. The accuracy of the linear representation of the network was already reported in \cite{Resener2016895}. The branch power losses are estimated by~\eqref{eq:pwl_model}. The loss in branch $i$~\eqref{eq:pwl_loss} depends on the branch resistance $R_{i}$ and the approximations $f_{i}^{\text{2Re}}$, $f_{i}^{\text{2Im}}$ of the squared real and imaginary currents, assuming $f_{i} = f_{i}^{\text{Re}} + jf_{i}^{\text{Im}}$. These approximations are defined by the tangent-line constraints~\eqref{eq:pwl_re} and~\eqref{eq:pwl_im}, where $a_{k}$ is the slope and $b_{k}$ the intercept, for all $k \in \{1, \dots, N^{\text{LC}}\}$ linear constraints.
\begin{subequations}
\label{eq:pwl_model}
\begin{align}
    & p_{i}^{\text{loss}} \approx R_{i} \left[ f_{i}^{\text{2Re}}
        + f_{i}^{\text{2Im}} \right]
        \label{eq:pwl_loss} \\
    & (f_{i}^{\text{Re}})^{2} \approx f_{i}^{\text{2Re}}
        \ge a_{k} f_{i}^{\text{Re}} + b_{k}
        \label{eq:pwl_re} \\
    & (f_{i}^{\text{Im}})^{2} \approx f_{i}^{\text{2Im}}
        \ge a_{k} f_{i}^{\text{Im}} + b_{k}
        \label{eq:pwl_im}
\end{align}
\end{subequations}

Each tangent line touches the parabola at a point $x_0$ and stays below it elsewhere, with slope $a_{k} = 2x_0$ and intercept $b_{k} = -x_0^2$. The $x_0$ points for the real and imaginary current components are spaced uniformly with $N^{\text{LC}}$ values over the intervals
$\{-f_{i}^{\text{Re,max}}, +f_{i}^{\text{Re,max}}\}$ and
$\{-f_{i}^{\text{Im,max}}, +f_{i}^{\text{Im,max}}\}$, respectively.

Kirchhoff's voltage and current laws are imposed
through~\eqref{eq:kirchhoff_model}. The voltage
law~\eqref{eq:kvl} relates branch currents to nodal voltage differences, where $\mathbf{S}$ is the node-branch incidence matrix, $\mathbf{V}$ the vector of
nodal voltage magnitudes, $[\mathbf{S}']_i \mathbf{V}$ the branch voltage differences, and $\Delta V_{m,t}$ the voltage contribution of an SVR where installed. The adjustment factor $K_{i,t}$ is adopted to improve the network representation accuracy, as developed in \cite{Resener2016895}. The current law is expressed for the real~\eqref{eq:kcl_re} and imaginary~\eqref{eq:kcl_im} parts, where $\mathbf{f}_{t}$ are the branch currents, $\mathbf{g}_{t}$ the current injections, $\mathbf{d}_{t}$ the load currents, and $\Delta \mathbf{f}_{t}$ the SVR current compensation upstream of the node; $\mathbf{q}_{t}^{\text{fcb}}$ and $\mathbf{q}_{t}^{\text{scb}}$ are the current injections of the fixed and SCB.
\begin{subequations}
\label{eq:kirchhoff_model}
\begin{align}
    & K_{i,t} R_{i}f_{i,t}^{\text{Re}} - X_{i}f_{i,t}^{\text{Im}}
        + [\mathbf{S}']_i \mathbf{V}_{t} - \Delta V_{m,t} = 0
        \label{eq:kvl} \\
    & \mathbf{S}\mathbf{f}_{t}^{\text{Re}} + \mathbf{g}_{t}^{\text{Re}}
        = \mathbf{d}_{t}^{\text{Re}} + \Delta \mathbf{f}_{t}^{\text{Re,SVR}}
        \label{eq:kcl_re} \\
    & -\mathbf{q}_{t}^{\text{fcb}} - \mathbf{q}_{t}^{\text{scb}}
        + \mathbf{S}\mathbf{f}_{t}^{\text{Im}} + \mathbf{g}_{t}^{\text{Im}}
        = \mathbf{d}_{t}^{\text{Im}} + \Delta \mathbf{f}_{t}^{\text{Im,SVR}}
        \label{eq:kcl_im}
\end{align}
\end{subequations}

\subsection{Building Thermal Dynamics Modeling} 

The resistance-capacitance (RC) building thermal dynamics model from \cite{Zhan2022364} represents SH for a single-zone building, formulated in~\eqref{eq:thermal_model}. Reduced-order building models serve as the prediction model in MPC for building control~\cite{Cole201469}, while the adopted single-resistance, single-capacitance (1R1C) model is commonly used to represent the dominant thermal behavior in building energy flexibility control methods~\cite{Demirel2025,Li2023}. 

The HP is commanded through the indoor-temperature setpoint $T_{e,t}^{\text{ind,set}}$: the indoor temperature is updated by the RC dynamics~\eqref{eq:temp_update} for all $t \in \{0,\ldots,T-1\}$, combining the heating contributions of the HP and the ERH, and the HP electrical power modulates so that the indoor temperature tracks the commanded setpoint~\eqref{eq:temp_track}. Here $\Delta t$ is the step length, $C^{\text{sh}}$ the thermal capacitance (kWh/K), $R^{\text{sh}}$ the thermal resistance (K/kW), $p_{t}^{\text{hp}}$ and $p_{t}^{\text{erh}}$ the HP and ERH power outputs (kW), $COP_{t}^{\text{sh}}$ and $COP^{\text{erh}}$ their respective COPs, and $T_{t}^{\text{out}}$ the outdoor temperature. The setpoint of each end-user $e$ is constrained to the thermal-comfort band~\eqref{eq:indoor_comfort}, between $T_{\text{min}}^{\text{ind}}$ and $T_{\text{max}}^{\text{ind}}$, and the indoor temperature inherits this band through~\eqref{eq:temp_track}. As a variable-speed unit, the HP is off or modulates between a minimum fraction $\gamma$ and its rated capacity~\eqref{eq:hp_modulation}, a semi-continuous behavior enforced by the binary $u_{e,t}^{\text{hp}}$; the ERH is switched in discrete stages $s \in \mathcal{S}$ of size $P^{\text{stage}}$~\eqref{eq:erh_stage}, selected by the binaries $u_{e,s,t}^{\text{erh}}$.
\begin{subequations} 
\label{eq:thermal_model} 
\begin{align}     
& \begin{aligned}
T_{t+1}^{\text{ind}} = T_{t}^{\text{ind}}
&+ \frac{\Delta t}{C^{\text{sh}}}
\left[ p_{t}^{\text{hp}} COP_{t}^{\text{sh}}
+ p_{t}^{\text{erh}} COP^{\text{erh}}  - \frac{1}{R^{\text{sh}}} \right.\\
&\left.
\left( T_{t}^{\text{ind}} - T_{t}^{\text{out}} \right) \right]
\end{aligned}
\label{eq:temp_update} \\     
& T_{e,t}^{\text{ind}} = T_{e,t}^{\text{ind,set}}         \label{eq:temp_track} \\     
& T_{\text{min}}^{\text{ind}} \leq T_{e,t}^{\text{ind,set}}         \leq T_{\text{max}}^{\text{ind}}         \label{eq:indoor_comfort} \\     
& \gamma\, P_{e,\text{hp}}^{\text{cap}}\, u_{e,t}^{\text{hp}}         \leq p_{e,t}^{\text{hp}}         \leq P_{e,\text{hp}}^{\text{cap}}\, u_{e,t}^{\text{hp}}         \label{eq:hp_modulation} \\     
& p_{e,t}^{\text{erh}} = \sum_{s \in \mathcal{S}}         P^{\text{stage}}\, u_{e,s,t}^{\text{erh}}         \label{eq:erh_stage}  
\end{align} 
\end{subequations}

The COP is the ratio of the heat output ($Q^{\text{hp}}$) to the electrical input power ($P^{\text{hp}}$). For the air-source HP under varying temperature, it is modeled by the quadratic regression~\eqref{eq:COP} \cite{Ruhnau2019}:
\begin{equation}
    COP = \alpha_c \left[ a_{0} + a_{1} \Delta T + a_{2} (\Delta T)^2 \right]
    \label{eq:COP}
\end{equation}
\noindent where $\alpha_c$ is a correction factor; $a_{0}$, $a_{1}$, $a_{2}$ are the constant, linear, and quadratic coefficients; and $\Delta T$ is the difference between the heat sink ($T_{t}^{\text{sink}}$) and heat source ($T_{t}^{\text{source}}$) temperatures at step $t$.

\subsection{Non-DLC Household Heating Simulation}

Households without DLC track a given target ($T_{e,t}^{\text{ref}}$) using a deadbeat thermostat simulated outside the optimization. The resulting HP and ERH powers are not decision variables but enter the network as loads, evaluated before each rolling solve. This approach covers non-DLC households in every operating regime. Inverting the thermal dynamics~\eqref{eq:temp_update} with the ERH off yields the setpoint-tracking HP power~\eqref{eq:db_req}, which is clamped to the admissible operating range~\eqref{eq:hp_modulation}. The resulting residual thermal deficit~\eqref{eq:db_def} is supplied by the minimum number of ERH stages required to meet it~\eqref{eq:db_erh}, with no ERH stage activated when the HP alone can meet the demand. The ERH is enabled only below the outdoor-temperature activation threshold $T^{\text{th}}$, corresponding to the HP balance temperature below which the HP alone cannot meet the heating demand. If the combined HP--ERH system cannot provide sufficient heating to reach the setpoint, the indoor temperature undershoots it according to~\eqref{eq:db_comfort}. 
\begin{subequations}
\label{eq:db_model}
\begin{align}
    & p_{e,t}^{\text{hp,req}} = \frac{1}{COP_{t}^{\text{sh}}}
        \!\left[ \frac{C^{\text{sh}}}{\Delta t}
        \big(T_{e,t+1}^{\text{ref}} - T_{e,t}^{\text{ind}}\big)
        + \frac{1}{R^{\text{sh}}}
        \big(T_{e,t}^{\text{ind}} - T_{t}^{\text{out}}\big) \right]
        \label{eq:db_req} \\
    & p_{e,t}^{\text{def}} = COP_{t}^{\text{sh}}
        \big(p_{e,t}^{\text{hp,req}} - p_{e,t}^{\text{hp}}\big)
        \label{eq:db_def} \\
    & p_{e,t}^{\text{erh}} = P^{\text{stage}}
        \min\!\Big( |\mathcal{S}|,\; \max\big(0,\; \big\lceil p_{e,t}^{\text{def}}/P^{\text{stage}} \big\rceil\big) \Big)
        \label{eq:db_erh} \\
    & T_{e,t}^{\text{ind}} \le T_{e,t}^{\text{ref}}
        \label{eq:db_comfort}
\end{align}
\end{subequations}

\subsection{Battery Energy Storage Systems }

The BESS charging and discharging operation is incorporated through the linear approximation adopted from \cite{Resener20231932}, formulated in~\eqref{eq:ess_model}. The state of charge at node $m$ and time $t$ ($SOC_{m,t}$) is bounded by its minimum and maximum limits~\eqref{eq:ess_soc_limits}, initialized in~\eqref{eq:ess_soc_init}, updated each step~\eqref{eq:ess_soc_update}, and closed cyclically over the day by \eqref{eq:ess_soc_cyclic} so that the end-of-day state of charge returns to its initial value, where $\eta_{m}^{\text{BESS}}$ and $\beta_{m}^{\text{BESS}}$ are the efficiency and self-discharge rate of the BESS at node $m$. The net real current injection~\eqref{eq:ess_current_net} is the difference between the non-negative charging and discharging currents~\eqref{eq:ess_current_nonneg}, whose mutually exclusive operating mode is enforced via big-$M$ with the binary $w_{m,t}$~\eqref{eq:ess_mode}, equal to $1$ in charging and $0$ in discharging mode. The maximum-current limits~\eqref{eq:ess_current_circle}, originally a quadratic feasible region, are approximated by supporting hyperplanes indexed by $k$~\eqref{eq:ess_current_poly} that inscribe the circle as a polygon.
\begin{subequations}
\label{eq:ess_model}
\begin{align}
    & SOC_{m}^{\text{min}} \leq SOC_{m,t} \leq SOC_{m}^{\text{max}}
        \label{eq:ess_soc_limits} \\
    & SOC_{m,0} = SOC^{\text{init}}
        \label{eq:ess_soc_init} \\
    & SOC_{m,t+1} = SOC_{m,t}
        + \Delta t \,\eta_{m}^{\text{BESS}} \!\left[ f_{m,t}^{\text{ch,Re,BESS}}
        - \tfrac{1}{\eta_{m}^{\text{BESS}}} f_{m,t}^{\text{dc,Re,BESS}} \right]
        \notag \\
    & \hphantom{SOC_{m,t+1} =\;} - \Delta t \,\beta_{m}^{\text{BESS}} SOC_{m,t}
        \label{eq:ess_soc_update} \\
    & SOC_{m,0} = SOC_{m,T} + \Delta t \,\eta_{m}^{\text{BESS}} \!\left[ f_{m,T}^{\text{ch,Re,BESS}}         - \tfrac{1}{\eta_{m}^{\text{BESS}}} f_{m,T}^{\text{dc,Re,BESS}} \right] \notag \\     
    & \hphantom{SOC_{m,0} =\;} - \Delta t \,\beta_{m}^{\text{BESS}} SOC_{m,T}         \label{eq:ess_soc_cyclic} \\
    & \mathbf{f}_{t}^{\text{Re,BESS}} = \mathbf{f}_{t}^{\text{ch,Re,BESS}}
        - \mathbf{f}_{t}^{\text{dc,Re,BESS}}
        \label{eq:ess_current_net} \\
    & \mathbf{f}_{t}^{\text{ch,Re,BESS}} \geq 0, \quad
      \mathbf{f}_{t}^{\text{dc,Re,BESS}} \geq 0
        \label{eq:ess_current_nonneg} \\
    & f_{m,t}^{\text{ch,BESS}} \le f_{m}^{\text{max,BESS}}\, w_{m,t}, \quad
      f_{m,t}^{\text{dc,BESS}} \le f_{m}^{\text{max,BESS}}\,(1 - w_{m,t})
        \label{eq:ess_mode} \\
    & \left( f_{m,t}^{\text{ch/dc,Re,BESS}} \right)^2
      + \left( f_{m,t}^{\text{Im,BESS}} \right)^2
      \leq \left( f_{m}^{\text{max,BESS}} \right)^2
        \label{eq:ess_current_circle} \\
    & f_{m,t}^{\text{ch/dc,Re,BESS}} \cos\theta_k
      + f_{m,t}^{\text{Im,BESS}} \sin\theta_k \le f_{m}^{\text{max,BESS}}
        \label{eq:ess_current_poly}
\end{align}
\end{subequations}

To limit the battery cycling, the number of charging and discharging mode changes over the horizon is bounded. A mode change at node $m$ and time $t$ occurs whenever the mode binary $w_{m,t}$ differs from $w_{m,t-1}$. This event is captured by the auxiliary switching indicator $\sigma_{m,t}^{\text{BESS}}$, which \eqref{eq:ess_sw_pos}--\eqref{eq:ess_sw_nn} constrain to satisfy $\sigma_{m,t}^{\text{BESS}} \ge |w_{m,t}-w_{m,t-1}|$ so that it equals $1$ on a switch and $0$ otherwise. The total number of switches over the horizon is then limited to $W_{m}^{\text{max,BESS}}$ by \eqref{eq:ess_sw_lim}.
\begin{subequations}
\label{ess:ess_switching_limits}
\begin{align}
    & \sigma_{m,t}^{\text{BESS}} \ge w_{m,t} - w_{m,t-1},         
        \quad \forall t \ge 1        
        \label{eq:ess_sw_pos} \\     
    & \sigma_{m,t}^{\text{BESS}} \ge w_{m,t-1} - w_{m,t},         
        \quad \forall t \ge 1        
        \label{eq:ess_sw_neg} \\     
    & \sigma_{m,t}^{\text{BESS}} \ge 0         
        \label{eq:ess_sw_nn} \\     
    & \sum_{t} \sigma_{m,t}^{\text{BESS}} \le W_{m}^{\text{max,BESS}}         
        \label{eq:ess_sw_lim}
\end{align}
\end{subequations}

\subsection{Switched Capacitor Banks}

The remotely dispatch of SCB is modeled as a discrete
selection of operating levels using a big-$M$ formulation with a constant impedance characteristic \cite{Resener2016895}, as presented
in~\eqref{eq:scb_model}. For each node $m$ and step $t$, the reactive current injection $q_{m,t}^{\text{scb}}$ is set to $l\,Y^{\text{ce}}V_{m,t}$ for the active level through the big-$M$ constraints~\eqref{eq:scb_lb}--\eqref{eq:scb_ub}, and exactly one level is selected by~\eqref{eq:scb_select}. Here $l \in \{0,\dots,C_{m}^{\text{scb}}\}$ are the discrete operating levels, $C_{m}^{\text{scb}}$ the number of available capacitor elements, $Y^{\text{ce}}$ the nominal admittance of a single element, $V_{m,t}$ the nodal voltage magnitude, $M$ a sufficiently large constant that relaxes the constraints for inactive levels, and $c_{m,l,t}$ a binary variable equal to $1$ if level $l$ is active and $0$ otherwise.
\begin{subequations}
\label{eq:scb_model}
\begin{align}
    & q_{m,t}^{\text{scb}} - l\,Y^{\text{ce}} V_{m,t}
        \geq -M\,(1 - c_{m,l,t})
        \label{eq:scb_lb} \\
    & q_{m,t}^{\text{scb}} - l\,Y^{\text{ce}} V_{m,t}
        \leq M\,(1 - c_{m,l,t})
        \label{eq:scb_ub} \\
    & \sum_{l \in C_{m}^{\text{scb}}} c_{m,l,t} = 1
        \label{eq:scb_select}
\end{align}
\end{subequations}

\subsection{Voltage-Regulation Devices}

The substation voltage magnitude ($V^{\text{ss}}_t$) is controlled by an OLTC,
formulated in~\eqref{eq:oltc_model}. The voltage is set by the integer tap
position $\tau_{t}^{\text{OLTC}}$~\eqref{eq:oltc_voltage}, where
$V^{\text{ss,min}}$ is the substation voltage at tap $0$ and $\Delta V ^{\text{OLTC}}$ is the voltage change per tap step. The tap position is encoded through the binary $\delta_{k,t}$~\eqref{eq:oltc_tap_position}, equal to $1$ if tap $k$ is selected, up to the maximum tap index $N_{\text{taps}}^{\text{OLTC}}$, with exactly one tap active at each time step enforced by~\eqref{eq:oltc_single_tap}.
\begin{subequations}
\label{eq:oltc_model}
\begin{align}
    & V^{\text{ss}}_t = V^{\text{ss,min}} + \Delta V ^{\text{OLTC}} \, \tau_{t}^{\text{OLTC}}
        \label{eq:oltc_voltage} \\
    & \tau_{t}^{\text{OLTC}} = \sum_{k=0}^{N_{\text{taps}}^{\text{OLTC}}}
        k \, \delta_{k,t}
        \label{eq:oltc_tap_position} \\
    & \sum_{k=0}^{N_{\text{taps}}^{\text{OLTC}}} \delta_{k,t} = 1
        \label{eq:oltc_single_tap}
\end{align}
\end{subequations}

The SVR is modeled as a type-B autotransformer equipped with LDC, adopting the linearized formulation from \cite{Ferraz2021}. For an SVR installed at node $m$ with downstream node $n$, it adjusts the integer tap position $\tau_{m,t}^{\text{SVR}}$ to maintain the voltage at the regulation node $n$ ($V_{n,t}$) within a bandwidth $\beta$ around the reference setpoint $V_{m}^{\text{set,SVR}}$. The LDC uses a potential transformer monitoring the local voltage $V_{m,t}$ and a current transformer monitoring the downstream branch currents $f_{mn,t}^{\text{Re}}$ and $f_{mn,t}^{\text{Im}}$, estimating $V_{n,t}$ from the voltage drop across the branch impedance.

The SVR effect enters Kirchhoff's voltage law~\eqref{eq:kvl} through
$\Delta V_{m,t}$~\eqref{eq:svr_dv} and the current laws~\eqref{eq:kcl_re},
\eqref{eq:kcl_im} through $\Delta f_{t}^{\text{Re,SVR}}$~\eqref{eq:svr_dfre} and
$\Delta f_{t}^{\text{Im,SVR}}$~\eqref{eq:svr_dfim} \cite{Resener2019378}, where $V_{k,t}$ is the upstream voltage. The voltage ratio
$a_{m,t}^{\text{SVR}}$~\eqref{eq:svr_ratio} depends on the tap step
$\Delta a_{m}^{\text{SVR}}$ (pu/tap), with $\tau_{m,t}^{\text{SVR}} \in \{\tau_{m}^{\text{SVR,min}},\ldots, \tau_{m}^{\text{SVR,max}}\}$ encoded in binary form~\eqref{eq:svr_binary}, where $K_{m}^{\text{SVR}}$ is the number of bits and $r_{h,m,t}$ a binary variable.
\begin{subequations}
\label{eq:svr_effect}
\begin{align}
    & \Delta V_{m,t} = V_{m,t} - V_{k,t}
        = \Delta a_{m}^{\text{SVR}} \tau_{m,t}^{\text{SVR}} V_{m,t}
        \label{eq:svr_dv} \\
    & \Delta f_{t}^{\text{Re,SVR}}
        = \Delta a_{m}^{\text{SVR}} \tau_{m,t}^{\text{SVR}} f_{mn,t}^{\text{Re}}
        \label{eq:svr_dfre} \\
    & \Delta f_{t}^{\text{Im,SVR}}
        = \Delta a_{m}^{\text{SVR}} \tau_{m,t}^{\text{SVR}} f_{mn,t}^{\text{Im}}
        \label{eq:svr_dfim} \\
    & a_{m,t}^{\text{SVR}} = 1 + \Delta a_{m}^{\text{SVR}} \tau_{m,t}^{\text{SVR}}
        \label{eq:svr_ratio} \\
    & \tau_{m,t}^{\text{SVR}} = \sum_{h=0}^{K_{m}^{\text{SVR}}}
        2^{h} r_{h,m,t} - \tau_{m}^{\text{SVR,max}}
        \label{eq:svr_binary}
\end{align}
\end{subequations}

Equations~\eqref{eq:svr_dv}, \eqref{eq:svr_dfre}, and~\eqref{eq:svr_dfim} are nonlinear due to the products of the integer tap $\tau_{m,t}^{\text{SVR}}$ with the voltage $V_{m,t}$ and the branch currents $f_{mn,t}^{\text{Re}}$, $f_{mn,t}^{\text{Im}}$. Substituting the binary encoding~\eqref{eq:svr_binary} and introducing auxiliary variables for the resulting bilinear terms yields an exact big-$M$ linearization, activated when an SVR is installed at node $m$ ($x_{m}^{\text{SVR}}=1$) and forcing the contributions to zero otherwise. The full set of disjunctive constraints can be found in \cite{Resener2019378}. 

The regulated voltage is held within the dead band~\eqref{eq:svr_band}, and a saturation relaxation~\eqref{eq:svr_sat_dn}--\eqref{eq:svr_sat_excl} permits the voltage to leave the band when the tap saturates, where $j_{m,t}^{\text{up}}$ and $j_{m,t}^{\text{down}}$ are binary saturation indicators, $B^{\text{max}}$ is the maximum binary tap sum, and $M$ a large constant.
\begin{subequations}
\label{eq:svr_regulation}
\begin{align}
    & V_{m}^{\text{set,SVR}} - \beta \leq V_{n,t}
        \leq V_{m}^{\text{set,SVR}} + \beta
        \label{eq:svr_band} \\
    & (V_{m}^{\text{set,SVR}} - \beta) - V_{n,t}
        \leq M j_{m,t}^{\text{down}}
        \label{eq:svr_sat_dn} \\
    & V_{n,t} - (V_{m}^{\text{set,SVR}} + \beta)
        \leq M  j_{m,t}^{\text{up}}
        \label{eq:svr_sat_up} \\
    & \sum_{h=0}^{K_{m}^{\text{SVR}}} 2^{h} r_{h,m,t}
        \geq B^{\text{max}} j_{m,t}^{\text{up}}
        \label{eq:svr_sat_link_up} \\
    & \sum_{h=0}^{K_{m}^{\text{SVR}}} 2^{h} r_{h,m,t}
        \leq B^{\text{max}} (1 - j_{m,t}^{\text{down}})
        \label{eq:svr_sat_link_dn} \\
    & j_{m,t}^{\text{up}} + j_{m,t}^{\text{down}} \leq 1
        \label{eq:svr_sat_excl}
\end{align}
\end{subequations}

\subsection{Control Regimes}

Control regimes are formulated for comparison, each changing a single element of the proposed rolling-horizon \textit{ST} regime, differing in the ST flexibility, the operation of the OLTC and SCB, the role of the ERH, or the BESS. The SVR voltage setpoint is optimized day-ahead in all cases. The control regimes are defined as follows:  
\begin{itemize}
    \item \textit{BASE}: the reference case that optimizes only the voltage regulation devices in a rolling-horizon MPC, with no DLC ($\Psi^{e}=\emptyset$); the ST setpoint variables and the thermal-comfort constraints~\eqref{eq:thermal_model} are not instantiated, and every household follows the non-DLC deadbeat thermostat~\eqref{eq:db_model} with a fixed setpoint, entering the network as an exogenous load.
    
    \item \textit{ST}: the proposed full rolling-horizon MPC that adds the co-optimized ST setpoint flexibility.
    
    \item \textit{ST$_{\text{VR-DA}}$}: as \textit{ST}, but the OLTC tap~\eqref{eq:oltc_tap_position} and SCB level~\eqref{eq:scb_select} are held at their committed day-ahead values, so only these devices lose their rolling-horizon recourse while the SVR still operates under LDC.  

    \item \textit{ST$_{\text{VR-local}}$}: as \textit{ST}, but the OLTC operates autonomously with LDC and the SCB is switched locally with a voltage-support voltage setpoint threshold, as presented Appendix~\ref{app:autonomous_OLTC_SCB}.  
    
    \item \textit{ST-DA}: as \textit{ST}, but the ST setpoints are committed to their Stage~1 day-ahead values and tracked by the deadbeat thermostat~\eqref{eq:db_model}, so the indoor temperature undershoots when the HP--ERH capacity cannot reach the setpoint~\eqref{eq:db_comfort}.  

    \item \textit{ST$_{\text{ERH-bkp}}$}: as \textit{ST}, but the ERH is restricted to a comfort backup that switches in stages only when the HP alone cannot hold the lower comfort bound, leaving preheating and any heating to reach indoor temperature targets above the comfort floor to the HP, as presented in Appendix~\ref{app:ST_ERH_bkp}.
    
    \item \textit{ST+BESS}: the proposed full model augmented with a co-optimized BESS~\eqref{eq:ess_model}--\eqref{ess:ess_switching_limits} bounded by an intra-day switching limit.  
\end{itemize}

The pairwise comparisons are designed to quantify the impact of the design choices of the proposed framework on network and heating-system operation. Comparing \textit{BASE} with \textit{ST} shows how much smart-thermostat flexibility alone relieves the network against the uncontrolled case. Comparing \textit{ST} with \textit{ST$_{\text{VR-DA}}$} reveals how much further co-scheduling the voltage-regulation devices within the rolling horizon lowers loading and lifts voltages. Comparing \textit{ST} with \textit{ST$_{\text{VR-local}}$} separates centralized VR co-optimization from the autonomous local response of the OLTC and the SCB. Comparing \textit{ST} with \textit{ST-DA} exposes what real-time setpoint recourse adds over a fixed day-ahead schedule. Comparing \textit{ST} with \textit{ST$_{\text{ERH-bkp}}$} shows the gain from dispatching the ERH as an active flexibility asset instead of holding it as an emergency comfort backstop. Comparing \textit{ST} with \textit{ST+BESS} shows the effect of dispatching a grid-scale BESS jointly with the ST setpoints, so that storage charging and discharging are scheduled together with preheating and setback. All regimes share the remaining network, thermal, and device constraints of the proposed model.

\section{Case Studies Setup}
\label{sec:case_study}

\begin{figure*}[pos=!htbp]
    \centering
        \begin{subfigure}{0.68\textwidth}
        \centering
        \includegraphics[width=\textwidth]{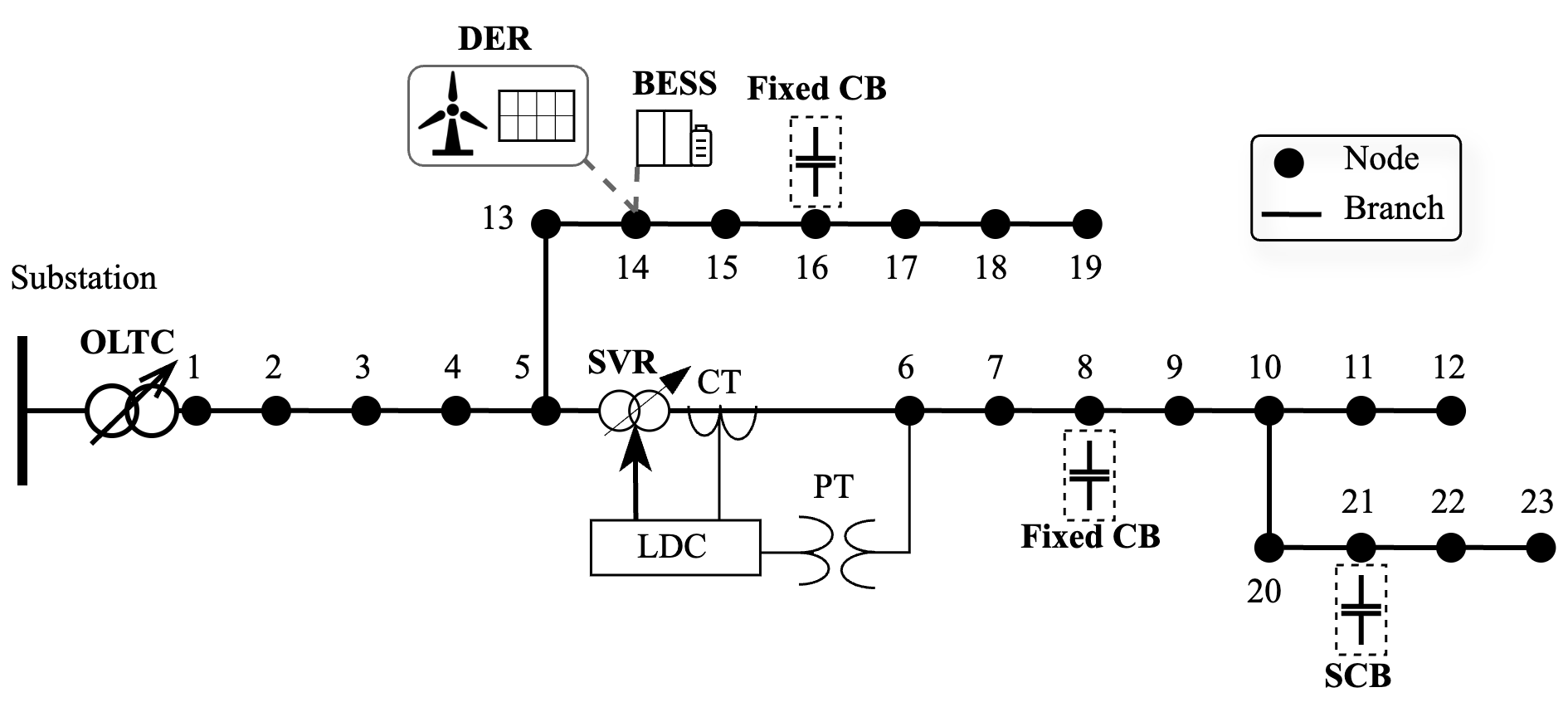}
        \caption{ }
        \label{fig:23node_dn}
        \end{subfigure}
    \begin{subfigure}{0.65\textwidth}
        \centering
        \includegraphics[width=\textwidth]{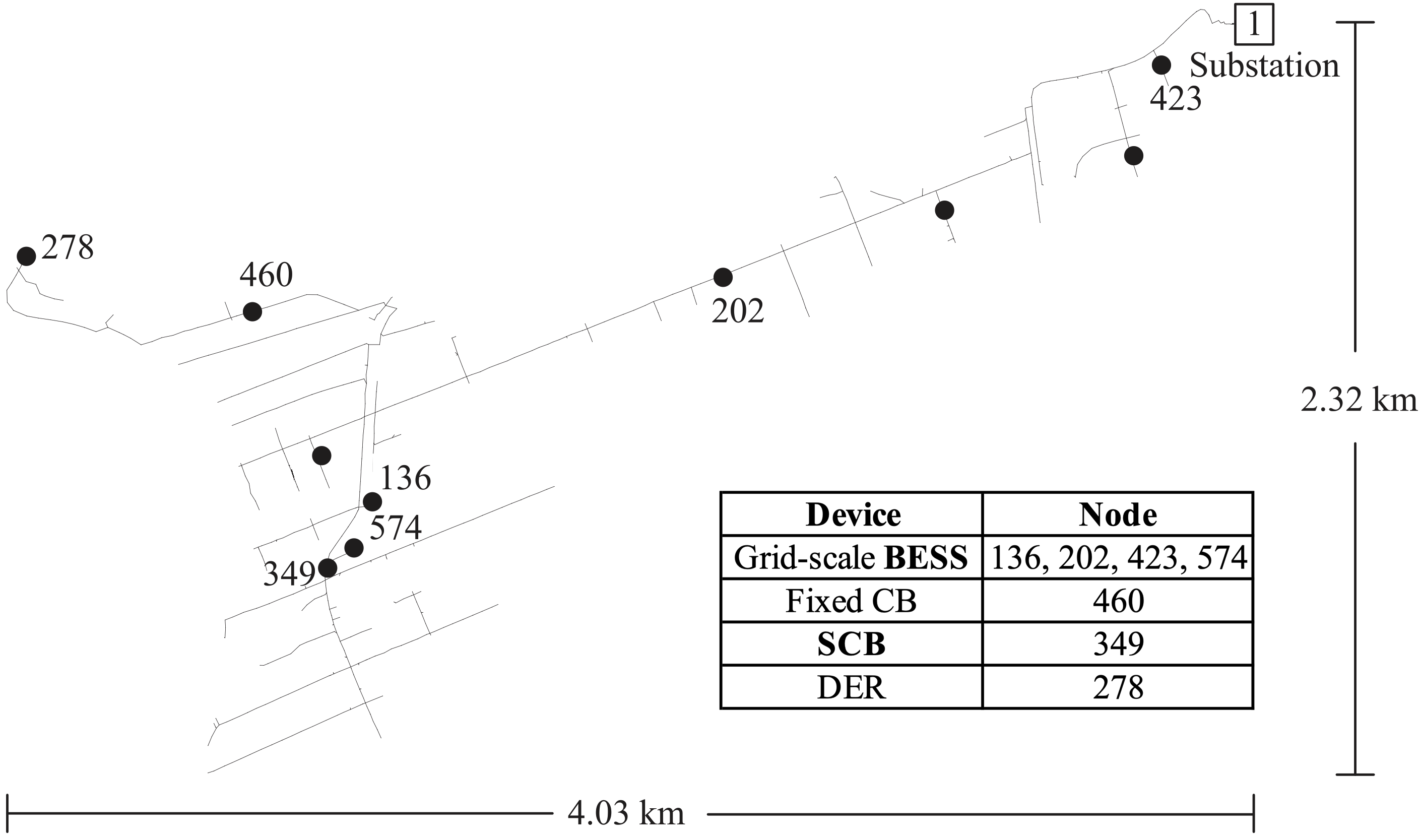}
        \caption{ }
        \label{fig:733node_dn}
    \end{subfigure}
    \caption{Distribution networks applied to the proposed MILP-based MPC model, including: (a) 23-node distribution network, adapted from \cite{Ferraz2021}; (b) 733-node distribution network, adapted from \cite{Resener20231932}. Additional details of the functioning of the LDC with the current transformer (CT) and potential transformer (PT) can be found in~\cite{Ferraz2021}.  }
    \label{fig:test_systems}
\end{figure*}

The proposed MILP-based MPC model is applied to the 23-node distribution network~\cite{Ferraz2021} and the 733-node real distribution network~\cite{Resener20231932}, shown in Fig.~\ref{fig:23node_dn} and Fig.~\ref{fig:733node_dn}. Both networks are three-phase and overhead, operating at a nominal voltage of 13.8~kV. It is assumed that the 23-node network supplies 1,760~end-users connected at transformers in nodes 2--23 over its 44~km extension (each branch 2~km long), with a maximum current of 419~A on branch 1--2. The 733-node network has 17.5~km of feeder length and supplies 3,215~end-users at 188~nodes, with a maximum allowed current of 525~A. Table~\ref{tab:device_specs} presents a summary of the specifications of the voltage regulation devices, BESS, and DER for both distribution networks. The 23-node network carries the full set of voltage regulation devices, including an SVR, whereas the 733-node network distributes storage across four BESS~\cite{Resener20231932}. The configuration of the autonomous operation of the OLTC and the SCB used for the ST$_{\text{VR-local}}$ control regime in the 23-node distribution network is presented in Appendix~\ref{app:autonomous_OLTC_SCB}. 

\begin{table}[htbp!]
\centering
\scriptsize
\setlength{\tabcolsep}{4pt}
\renewcommand{\arraystretch}{1.2}
\caption{Voltage-regulation device, BESS, and DER specifications for the 23-node and 733-node distribution networks.}
\label{tab:device_specs}
\begin{tabularx}{\linewidth}{@{}>{\raggedright\arraybackslash}X >{\raggedright\arraybackslash}p{0.46\linewidth}@{}}
\toprule
Parameter & Value \\
\midrule
\multicolumn{2}{@{}l}{\textit{Substation on-load tap changer }}\\
Minimum substation voltage $V^{\text{ss,min}}$ & 1.00~pu \\
Tap step $\Delta V^{\text{OLTC}}$ & 0.0125~pu \\
Maximum tap $N_{\text{taps}}^{\text{OLTC}}$ & 4 (max.\ $V_{t}^{\text{ss}}=1.05$~pu) \\
\midrule
\multicolumn{2}{@{}l}{\textit{Fixed capacitor banks}}\\
Rating & 600~kvar \\
Location (node) & 8, 16 (23-node); 460 (733-node) \\
\midrule
\multicolumn{2}{@{}l}{\textit{Switched capacitor banks}}\\
Modules & $2\times600$~kvar \\
Location (node) & 21 (23-node); 349 (733-node) \\
\midrule
\multicolumn{2}{@{}l}{\textit{Step voltage regulator, 23-node only~\cite{Ferraz2021}}}\\
Location / regulated node & 5 / 6 (LDC) \\
Regulation range & $\pm10\%$ (0.10~pu) \\
Tap positions $\tau_{m,t}^{\text{SVR}}$ & $\{-16,\dots,16\}$ \\
Tap step $\Delta a_{m}^{\text{SVR}}$ & 0.00625~pu \\
Voltage ratio $a_{m,t}^{\text{SVR}}$ & $[0.9,\,1.1]$~pu \\
Maximum current & 400~A \\
Deadband $\beta$ & 0.01 \\
\midrule
\multicolumn{2}{@{}l}{\textit{Battery energy storage systems}}\\
Rating & 0.8~MW / 0.8~MWh (1C-rate)~\cite{Resener20231932} \\
\multirow{2}{*}{Location (node)} & 14 (23-node); 136, 202, 423, 574 (733-node)~\cite{Resener20231932} \\
Efficiency $\eta_{m}^{\text{BESS}}$ & 0.95 \\
Self-discharge $\beta_{m}^{\text{BESS}}$ & $4 \times 10^{-5}$ \\
SOC limits & $[0.1,\,0.9]\,SOC_{\text{nom}}$ \\
Initial SOC & $0.5\,SOC_{\text{nom}}$ \\
Maximum mode changes $W_{m}^{\text{max,BESS}}$ & 6 \\
\midrule
\multicolumn{2}{@{}l}{\textit{Distributed energy resource}}\\
Location (node) & 14 (23-node); 278 (733-node)~\cite{Ferraz2021} \\
\multirow{2}{*}{Generation} &  Nominal (1.0~MW); peak (0.84~MW)~\cite{Ferraz2021} \\ 
\bottomrule
\end{tabularx}
\end{table}

\subsection{Building Thermal Dynamics Characteristics and Weather Data}

Table~\ref{tab:building_heating} presents a summary of the specifications of the building thermal model and heating devices. Each house is represented by a first-order RC thermal model whose parameters span the range of insulation and thermal-mass levels found in the Canadian residential stock~\cite{Swan200975}, with reference values corresponding to a moderately insulated 1980s dwelling~\cite{Zhan2022364,aschidamini2026122015}. To introduce physical heterogeneity across the network, the thermal resistance $R^{\text{sh}}$ and capacitance $C^{\text{sh}}$ are sampled independently from the uniform ranges. Over this envelope, the resulting per-house SH demand, the delivered heat required to maintain the indoor setpoint across the 2014--2024 winter record, spans approximately $6.2$ to $9.8$~MWh per winter, from the most to the least insulated dwelling. 

\begin{table}[htbp!] 
\centering 
\scriptsize 
\setlength{\tabcolsep}{4pt} 
\renewcommand{\arraystretch}{1.15} 
\caption{Building thermal model and heating-device specifications.} 
\label{tab:building_heating} 
\begin{tabularx}{7.5cm}{@{}
    >{\hsize=0.5\hsize\raggedright\arraybackslash}X
    >{\hsize=0.5\hsize\raggedright\arraybackslash}X
@{}} 
\toprule 
Parameter & Value \\ 
\midrule 
\multicolumn{2}{@{}l}{\textit{Building thermal model}}\\ 
Reference floor area & 149.2~m$^2$~\cite{Swan200975}   \\ 
Thermal resistance $R^{\text{sh}}$  & 4.44; $(3.44,5.44)$~K/kW~\cite{Zhan2022364}   \\ 
Thermal capacitance $C^{\text{sh}}$  & 20; $(15,25)$~kWh/K~\cite{Zhan2022364}   \\ 
Per-house SH demand & 6.2 to 9.8~MWh/winter  \\ 
\midrule 
\multicolumn{2}{@{}l}{\textit{Air-source heat pump}} \\
Rated capacity catalog & $\{3.0,3.5,4.0\}$~kW    \\ 
Modulation range $\gamma$ & 20 to 100\%   \\ 
COP coefficients $a_0,a_1,a_2$ & $6.08,\,-0.09,\,0.0005$~\cite{Zhan2022364,Ruhnau2019}   \\ COP correction factor $\alpha_c$ ~\cite{Ruhnau2019} & 0.85   \\ 
Displacement power factor & 0.98 lagging   \\ 
\midrule 
\multicolumn{2}{@{}l}{\textit{Backup electric resistance heater }}\\ 
Capacity $P^{\text{cap}}_{\text{erh}}$ & 10~kW   \\ 
Staging & $4\times2.5$~kW    \\ 
Balance-point (activation) temp. & $-13.4$ to $-9.4$\,\textdegree C (median $-11.7$)   \\ 
\bottomrule 
\end{tabularx} 
\end{table}

Each air-source HP is sized to its building envelope at a design outdoor temperature of $-12$\textdegree C. Moreover, historical outdoor temperature data from \cite{weather_data_vancouver_airport} for Richmond, Canada for the Winter months (December--February) from 2014 to 2024 were used. The design outdoor temperature is deliberately colder than the conventional design temperature for the site, the 1st percentile of the winter record is $-7.4$\textdegree C, and is instead anchored to the historical coldest scenario that drives HP--ERH coordination: it lies between the average ($-10.9$\textdegree C) and minimum ($-14.5$\textdegree C), and only $0.13\%$ of the winter hours in the record fall below it. For every house, the required HP electrical input is obtained by taking the building's steady-state heat loss, the difference between the $20$\textdegree C indoor setpoint and the $-12$\textdegree C design temperature, divided by the thermal resistance $R^{\text{sh}}$, and dividing it by the HP COP at that temperature. The result is rounded to the nearest entry in a discrete catalog of commercial ratings. Capacity therefore scales inversely with insulation, since leakier houses (lower $R^{\text{sh}}$) have higher heat loss and receive larger units. The HP COP follows the temperature-dependent air-source quadratic of~\cite{Ruhnau2019} and is assumed independent of the modulation level.

Each house has a backup ERH with activation threshold equal to the balance-point temperature, the outdoor temperature at which the building's steady-state heat loss equals the maximum thermal output of its assigned HP. Because HP capacity is matched to each envelope, these balance points cluster near the design temperature rather than spanning a wide range (between $-9.4$\textdegree C and $-13.4$\textdegree C). Consequently, for the historical coldest scenario, the ERH engages only during the coldest hours when temperatures fall below $-12$\textdegree C for $29\%$ of the day, and remains off otherwise, when temperatures stay above the thresholds.  

The temperature deviation index ($T_{\text{dev}}$) and the deviation period index ($P_{\text{dev}}$), presented in \eqref{eq:temperature_deviation_index} and \eqref{eq:deviation_period_index} (adopted from \cite{Esrafilian-Najafabadi2022,aschidamini2026122015}), are used to assess indoor comfort. The temperature deviation index, considering only the indoor temperature, is defined as the mean absolute error of the indoor temperature over a predefined time horizon, whereas the deviation period index represents the percentage of time during which the indoor temperature deviation exceeds 0.5\textdegree C. 
\begin{align}
    T^{\text{dev}} &= \frac{1}{T}\sum_{t=1}^{T}\left|T_{t}^{\text{ind}} - T^{\text{ind,set}}\right| \label{eq:temperature_deviation_index}\\
    P^{\text{dev}} &= \frac{\sum_{t=1}^{T} C^{\text{dev}}_{t}}{T} \times 100 \label{eq:deviation_period_index}
\end{align}
where $C^{\text{dev}}_{t} = 1$ if $\left|T_{t}^{\text{ind}} - T^{\text{ind,set}}\right| > 0.5$\textdegree C, and $C^{\text{dev}}_{t} = 0$ otherwise.

\subsection{Control Configurations}

The MPC prediction look-ahead is $T=8$ time steps of length $\Delta t =0.5$~h, with indoor temperature bounds $T_{\text{min}}^{\text{ind}}=19$\textdegree C and
$T_{\text{max}}^{\text{ind}}=21$\textdegree C. The uncontrolled houses operate with a fixed indoor-temperature setpoint $T^{\text{ind,set}} = 20\,^{\circ}$C. Initial indoor temperatures are drawn from $20\pm0.5$~°C. The cost of energy losses is $C^{\text{kWh}}=0.11~\$/\text{kWh}$, corresponding to the retail electricity price, while the cost of voltage violations is $C^{\text{v}}=100~\$/\text{pu}$. Bounds of $V^{\text{min}}=0.95$~pu and $V^{\text{max}}=1.05$~pu are considered per ANSI C84.1 \cite{ANSI_C84_1_2020}. The MILP-based MPC model was implemented in Pyomo \cite{bynum2021pyomo} and solved with Gurobi \cite{gurobi} through its Python \cite{python} interface, using a relative mixed-integer programming gap tolerance of $1\%$.

The proposed framework and baselines are evaluated across scenarios that vary the HP penetration level and the DLC rate of ST, each assessed over a set of realized base-load days and reported as mean $\pm$ standard deviation. The HP penetration level is the share of houses with HP, with the remainder using gas heaters, whereas the DLC rate is the share of HP and ERH houses granting DLC setpoint control. To capture day-to-day demand variability rather than a single representative day, a total of 5 and 15 realized inflexible residential base-load days are used for the 23-node and 733-node networks, respectively. The hourly active and reactive nodal demand data, which exclude electric heating loads, are obtained from~\cite{Ferraz2021} and provide, for each network, both the realized base-load days and a single representative day used as the day-ahead forecast. Against a predicted day-ahead peak of 5.02~MW (23-node) and 4.44~MW (733-node), the realized feeder peak spans 5.35--6.04~MW and 4.17--5.77~MW, a day-to-day coefficient of variation of 5.4\% and 8.0\%, respectively, so the realized days consistently exceed the day-ahead prediction, and the rolling-horizon stage must regulate under higher, more variable peaks than anticipated.

\section{Numerical Results }
\label{sec:results}

The coordinated framework (regimes ST and ST+BESS) is compared against a no-DLC regime (BASE), a day-ahead demand-side schedule (ST-DA), a restriction of the ERH to
a thermal-comfort backstop (ST$_{\text{ERH-bkp}}$), an autonomous local control of the voltage-regulation devices (ST$_{\text{VR-local}}$), and a day-ahead commitment of the voltage regulation devices (ST$_{\text{VR-DA}}$). Performance is reported through heating energy use, feeder peak demand, loss and voltage violation cost, nodal voltage, asset loading, and indoor thermal comfort.

\subsection{Application Example: 23-Node Distribution Network}

The 23-node network has considerable network length (2~km per section) and high system loading~\cite{Resener2016895,Resener20231932}, making it particularly suitable for evaluating the proposed framework under voltage-constrained operating conditions. We quantify the value of ST and BESS flexibility and, subsequently, illustrate the proposed rolling co-optimization of demand-side flexibility, voltage regulation assets, and BESS.

\subsubsection{Value of Demand-Side Flexibility}

\begin{figure}[pos=htbp!]
    \centering
    \includegraphics[width=1.0\linewidth]{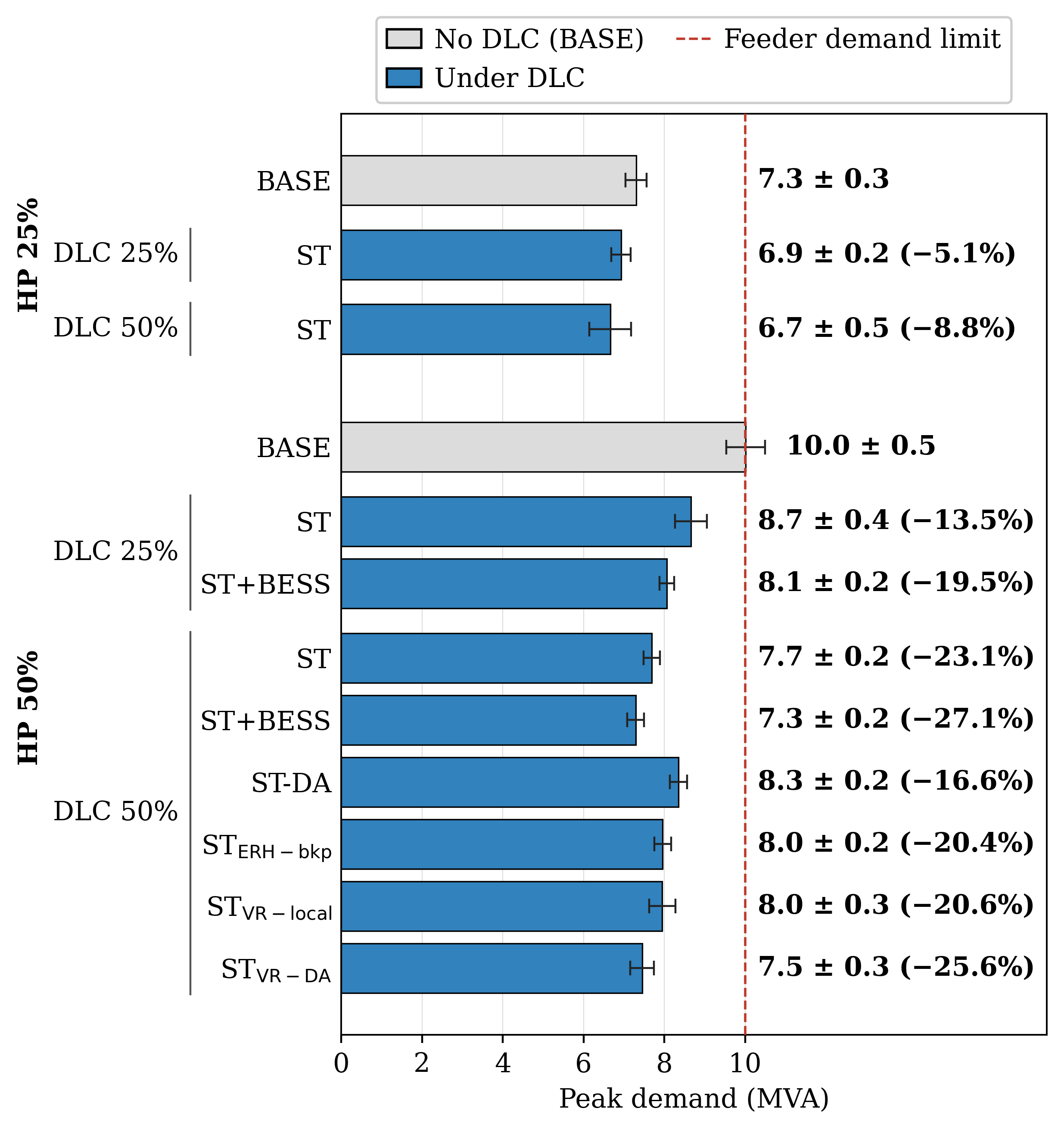}
        \caption{Peak demand across control regimes for the 23-node distribution network.}
    \label{fig:peak_demand_23node}
\end{figure}

Fig.~\ref{fig:peak_demand_23node} presents the resulting peak demand across control regimes under HP penetration levels and DLC rates, while Table~\ref{tab:results_TS2_combined_23node} reports the objective function (OF) terms across the same control regimes. At 50\% HP penetration, the control regime ST reduces the peak to 8.7~MVA at a 25\% DLC rate and to 7.7~MVA at 50\% DLC rate, reductions of 13.5\% and 23.1\%, respectively. The BESS shaves another approximately 0.6~MVA and 0.4~MVA, bringing the peak to 7.3~MVA, a 27.1\% reduction. 
The peak shaving of the proposed framework is achieved together with the lowest system cost, as ST and ST+BESS regimes reach an average objective-function cost of \$818.1--827.1 at 50\% DLC rate, a 17\% reduction from BASE, while cutting the voltage-violation cost from \$19.0 (BASE) to \$1.3--1.5. Although ST$_{\text{VR-DA}}$ attains 7.5~MVA peak, its day-ahead voltage setpoints raise the voltage-violation cost to \$85.8, above even the BASE case, so its total cost falls by only 7\%. The proposed framework therefore delivers comparable peak reduction while nearly eliminating voltage-violations costs. These benefits arise from co-optimizing the voltage regulation devices with the demand-side flexibility of the ST and the BESS, which displaces demand away from the feeder peak while jointly minimizing the costs of power losses and voltage violations.
\begin{table*}[pos=htbp!]
\centering
\setlength{\tabcolsep}{2pt}
\scriptsize
\caption{Objective function terms for the proposed framework under control regime ST and ST+BESS, for BASE (no-DLC), and remaining control regimes for the 23-node distribution network. }
\label{tab:results_TS2_combined_23node}
\begin{tabularx}{12cm}{@{}
    >{\hsize=0.30\hsize\raggedright\arraybackslash}X
    >{\hsize=0.25\hsize\raggedright\arraybackslash}X
    >{\hsize=0.20\hsize\raggedright\arraybackslash}X
    >{\hsize=0.20\hsize\centering\arraybackslash}X
    >{\hsize=0.20\hsize\centering\arraybackslash}X
    >{\hsize=0.20\hsize\centering\arraybackslash}X
@{}} 
\toprule
 HP penetration (\%) &
 Control regime & DLC rate (\%) &
 OF (\$) &
 $c^{\text{loss}}$ (\$) &
 $c^{\text{volt}}$ (\$) \\
\midrule
0 & BASE & ---  & $203.3 \pm 28.8$ & $203.2 \pm 28.8$ & $0.1 \pm 0.1$ \\
\midrule
\multirow{3}{*}{25}
 & BASE & --- & $499.6 \pm 43.6$ & $499.1 \pm 43.7$ & $0.4 \pm 0.1$ \\
 & ST & 25   & $478.9 \pm 42.0$ & $477.7 \pm 42.1$ & $1.3 \pm 0.8$ \\
 & ST & 50  & $451.2 \pm 40.5$ & $450.4 \pm 40.3$ & $0.8 \pm 0.3$ \\
\midrule
\multirow{9}{*}{50}
 & BASE & ---   & $992.9 \pm 71.6$ & $973.8 \pm 60.4$ & $19.0 \pm 13.3$ \\
 & ST & 25    & $910.2 \pm 67.6$ & $904.4 \pm 65.4$ & $5.8 \pm 2.7$ \\
 & ST+BESS & 25   & $896.2 \pm 62.1$ & $893.5 \pm 61.1$ & $2.7 \pm 1.5$ \\

 & ST & 50  & $827.1 \pm 57.3$ & $825.6 \pm 57.6$ & $1.5 \pm 0.8$ \\

 & ST+BESS  & 50  & $818.1 \pm 56.7$ & $816.8 \pm 56.3$ & $1.3 \pm 0.7$ \\
 
 & ST-DA & 50  & $848.8 \pm 60.8$ & $847.0 \pm 59.1$ &  $1.8 \pm 2.1$ \\

 & ST$_{\text{ERH-bkp}}$ & 50  & $836.6 \pm 59.4$ & $833.5 \pm 58.6$ & $3.1 \pm 1.9$ \\

 & ST$_{\text{VR-local}}$ & 50  & $843.9 \pm 60.4$ & $842.6 \pm 59.1$ & $1.3 \pm 1.6$ \\
 
 & ST$_{\text{VR-DA}}$ & 50  & $925.3 \pm 111.8$ & $839.5 \pm 63.1$ & $85.8 \pm 48.8$ \\ 
\bottomrule
\end{tabularx}
\end{table*}

\begin{figure*}[htbp!]
    \centering

    \begin{subfigure}[t]{0.25\textwidth}
        \centering
        \includegraphics[width=\linewidth]{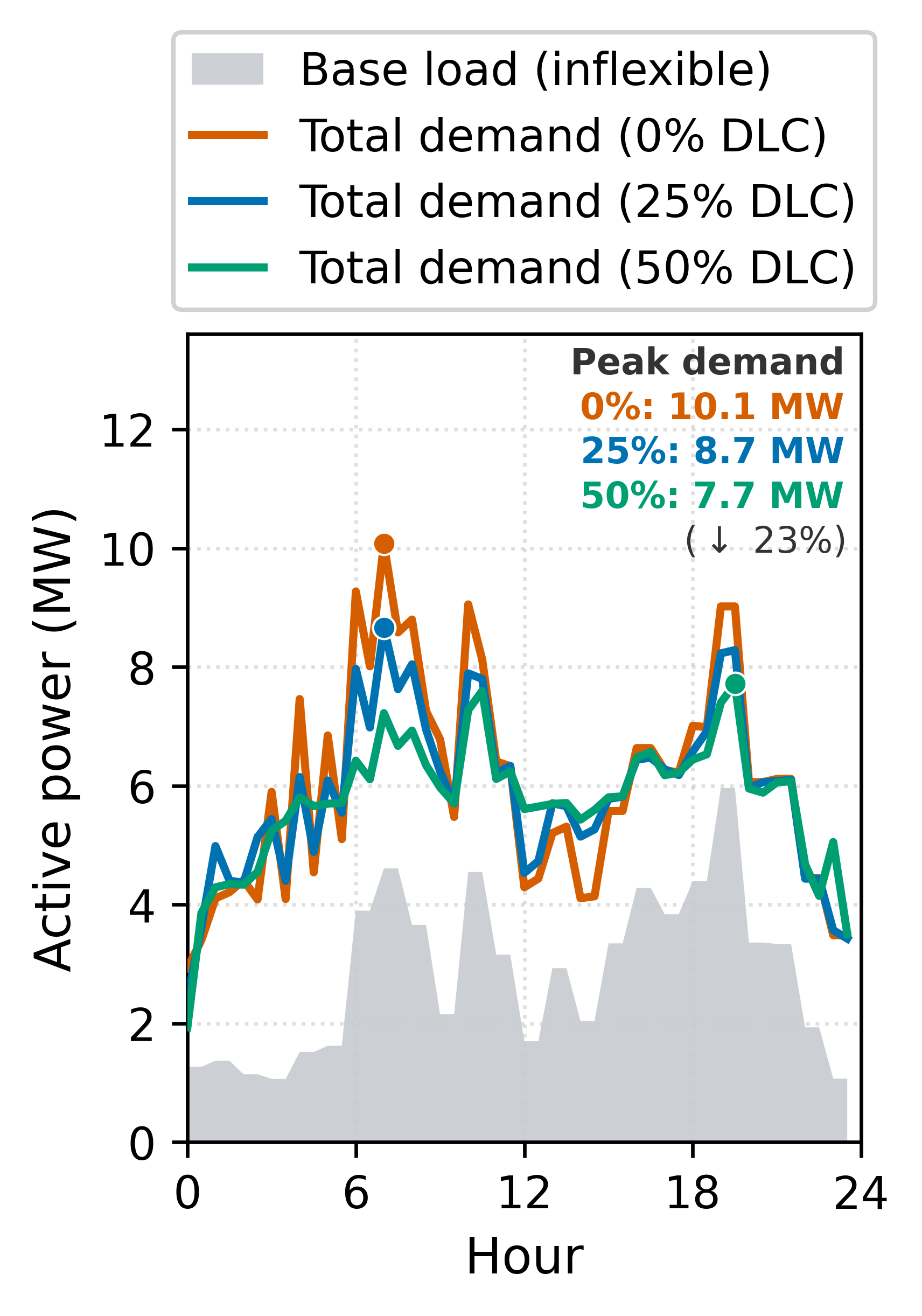}
        \caption{}
        \label{fig:total_vs_base_HP50_23node}
    \end{subfigure}
    \hfill
    \begin{subfigure}[t]{0.70\textwidth}
        \centering
        \includegraphics[width=\linewidth]{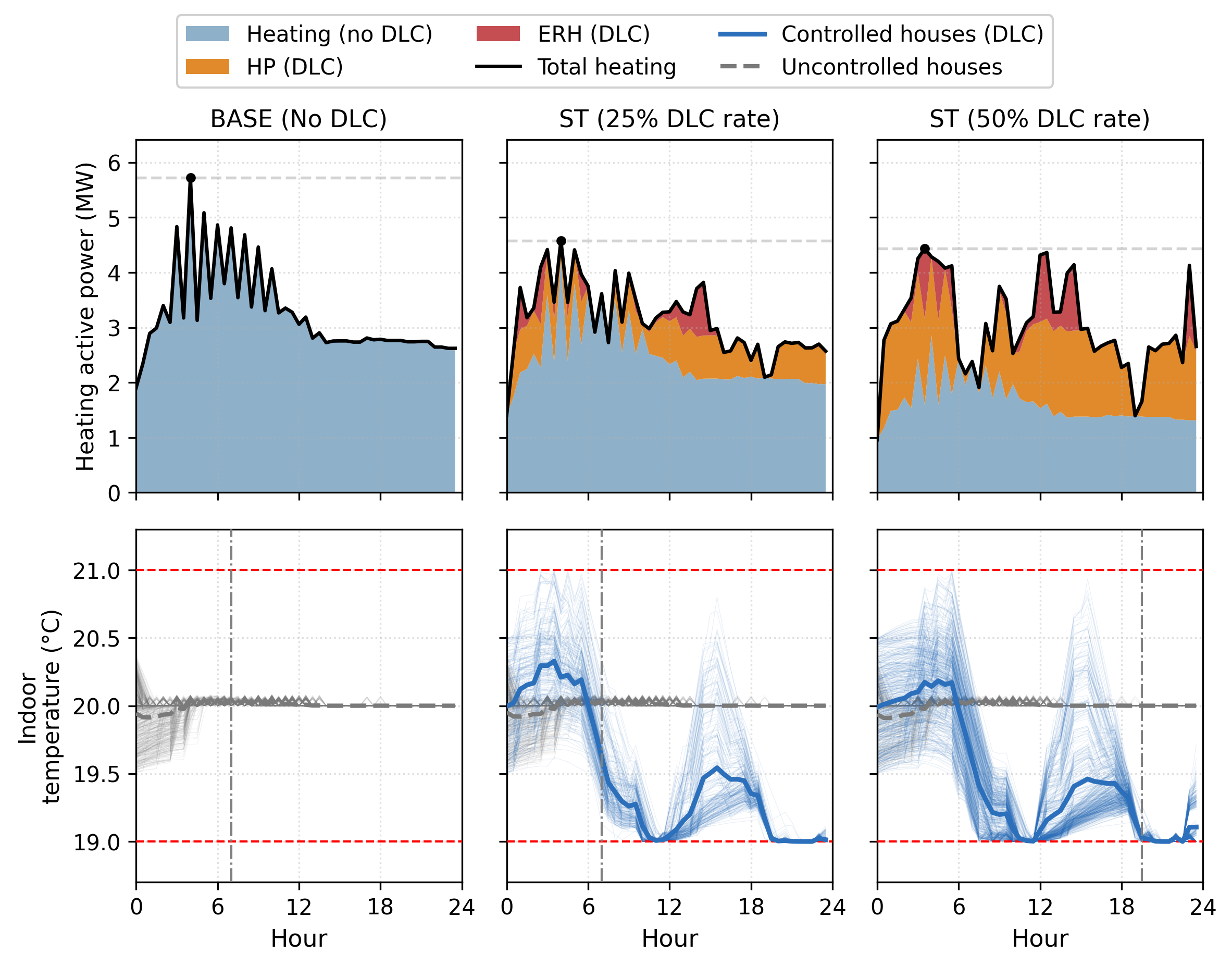}
        \caption{}
        \label{fig:heating_load_mpc_HP50_23node}
    \end{subfigure}

    \caption{Comparison of the proposed MPC operation under different levels of ST demand-side flexibility for 50\% HP penetration in the 23-node distribution network: 
    (a) total active power demand over time; 
    (b) heating demand composition and indoor temperature trajectories for uncontrolled and controlled (DLC) houses, with a vertical line sitting at the feeder peak time. }
    \label{fig:active_power_demand_HP50_23node}
\end{figure*}

Fig.~\ref{fig:active_power_demand_HP50_23node} reports, for a single representative base-load scenario at 50\% HP penetration, how the ST control regime shaves the feeder peak by reshaping the HP and ERH heating demand. The feeder demand peaks at 7~AM, as shown in Fig.~\ref{fig:total_vs_base_HP50_23node}, and ST control cuts this peak by 13\% at a 25\% DLC rate and by 23\% at a 50\% DLC rate. As shown in Fig.~\ref{fig:heating_load_mpc_HP50_23node}, the controlled buildings are preheated before the peak by the HP together with the ERH, which is used for preheating rather than only as a fallback for comfort, since the COP of the air-source HP drops significantly in colder weather, exactly when the feeder peak tends to appear on feeders with high heating electrification. The feeder peak remains at 7~AM, but the setback drives the indoor temperature toward the lower bound of 19\textdegree C, rather than maintaining the 20\textdegree C setpoint through variable-speed HP modulation.

Re-optimizing the ST setpoints over the rolling horizon, rather than committing them day-ahead as in ST-DA, sharpens the peak reduction. At 50\% HP penetration with a 50\% DLC rate, the rolling-horizon ST lowers the feeder peak to 7.7~MVA against 8.3~MVA for ST-DA, a reduction of 23.1\% against 16.6\% relative to BASE, as shown in Fig.~\ref{fig:peak_demand_23node}. It also lowers the objective-function cost to \$827.1 against \$848.8, as reported in Table~\ref{tab:results_TS2_combined_23node}, because the controller observes the realized demand and re-dispatches the preheating and setback signals accordingly.

\begin{figure}[pos=htbp!]
    \centering
    \includegraphics[width=1.0\linewidth]{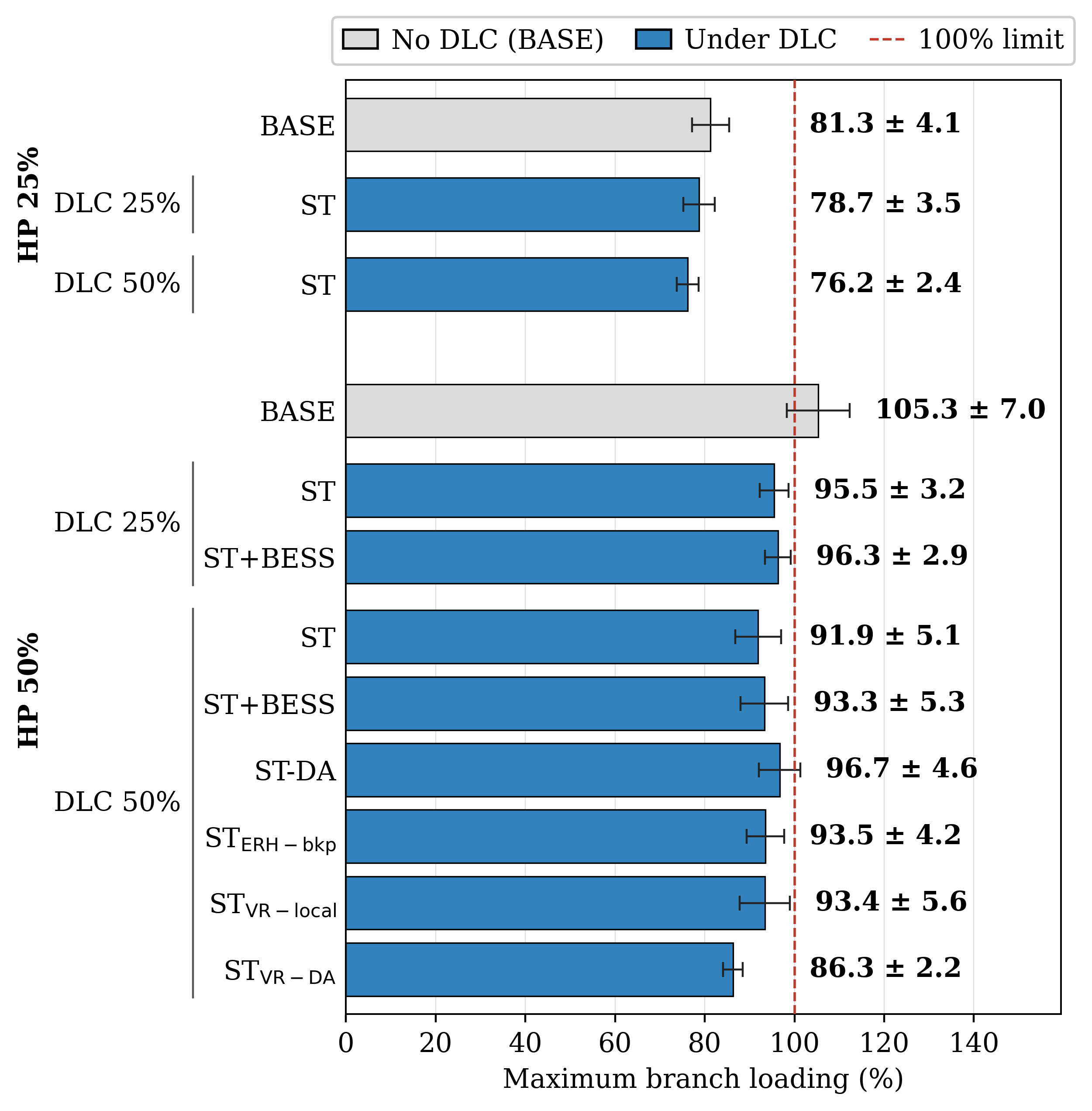}
        \caption{Maximum branch loading across control regimes for the 23-node distribution network.}
    \label{fig:branch_loading_23node}
\end{figure}

Fig.~\ref{fig:branch_loading_23node} presents the maximum branch loading across control regimes. The proposed ST control regime decreases the maximum branch loading for the 23-node distribution network, which does not have significant thermal issues, as only one branch has overloading up to 50\% HP penetration. At 50\% HP penetration, the DLC rate at 25\% clears the overloaded branches, which under BASE control regime stay at 105.3~$\pm$~7.0\%. The control regime ST$_{\text{VR-DA}}$ has a lower maximum branch loading at 86.2\%, which is also within the limits. 

\subsubsection{Heating Energy Use, Comfort, and Efficiency}

\begin{figure}[pos=htbp!]
    \centering
    \includegraphics[width=1.0\linewidth]{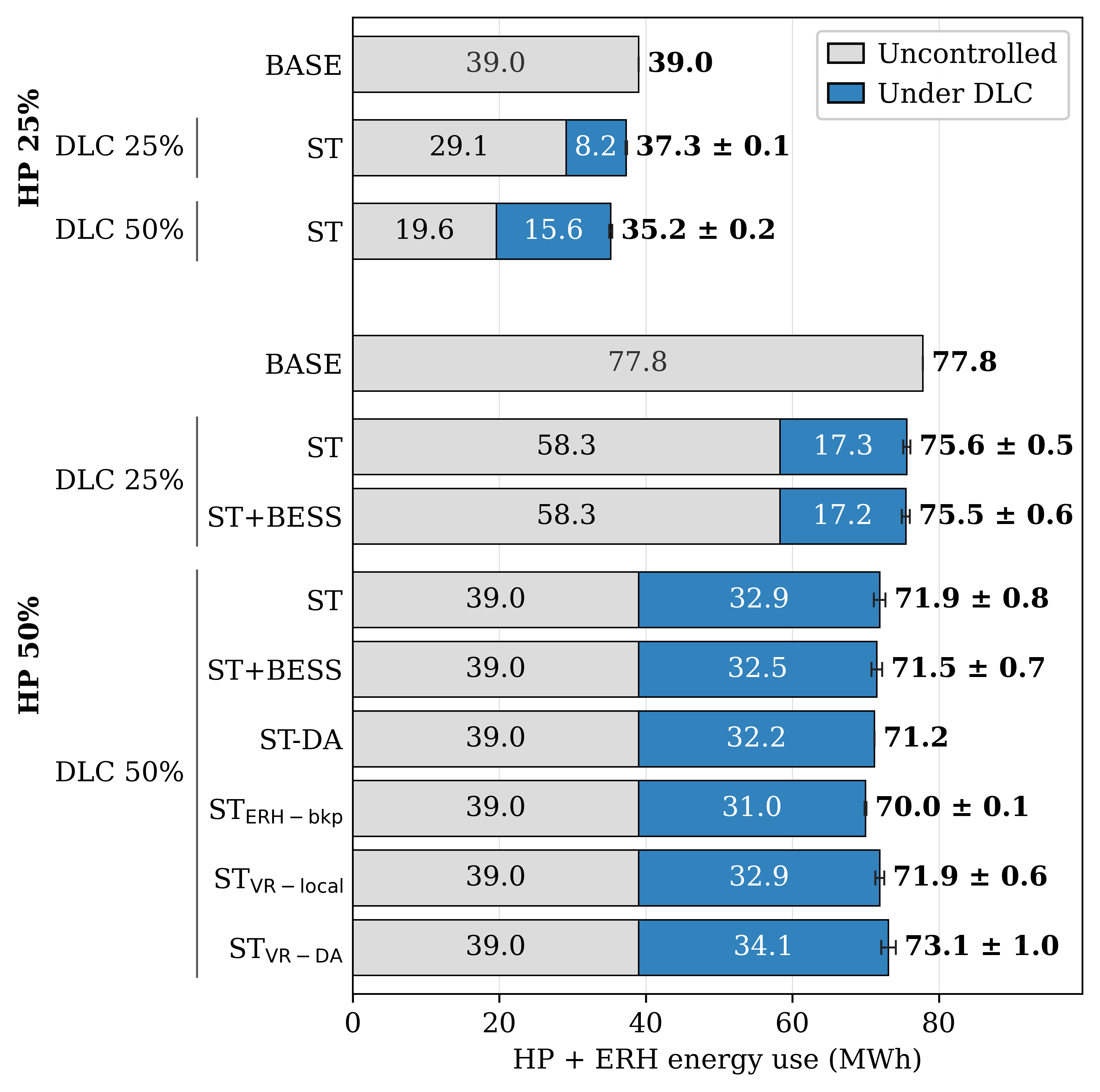}
        \caption{Energy use of HP+ERH across control regimes for the 23-node distribution network.}
    \label{fig:energy_use_hp_erh_23node}
\end{figure}

Fig.~\ref{fig:energy_use_hp_erh_23node} presents the total heating energy use and its division into dispatchable and non-dispatchable components. The results show that DLC of the ST not only shifts heating energy use into the dispatchable category but also reduces the total heating electricity consumption, by about 8\% at 50\% HP penetration (from 77.8~MWh to 71.5~MWh for ST+BESS) and by up to 10\% at 25\% HP penetration. This reduction results from temperature setbacks while indoor temperatures remain within the comfort band, indicating that preheating within this band does not increase the overall heating energy use. 

Table~\ref{tab:comfort_TS2_23node} presents the energy use reduction, temperature deviation index, and deviation period indexes over the houses and base-load realizations across control regimes. At the household-level, the feeder-wide energy use reduction is not uniform across buildings. The ST control regime reduces the average energy use, measured against the expected uncontrolled heating of the same house under the realized weather. The reduction persists across control regimes and HP penetration level ($-21.3$\% to $-11.8$\% on average). At 50\% HP penetration, however, the ST control raises the energy use of a minority of the controlled buildings, 17.3\% at 25\% DLC rate and 5.0\% at 50\% DLC rate, with the largest single-building increase in electricity cost on any day being \$1.74 and \$2.02 (at a retail price of \$0.11/kWh), respectively. These increases arise where the controller shifts heating earlier (preheating within the comfort band) and relies more on lower-COP ERH when HP capacity is insufficient during the coldest periods, yet the additional consumption is less than 1\% of the total baseline heating energy. Because the model optimizes at the feeder level and does not compensate individual households, the design of monetary incentives for the DR program should account for the buildings whose energy use rises. 
\begin{table*}[htbp!]
\centering
\setlength{\tabcolsep}{2pt}
\scriptsize
\caption{Per-household effects of the DLC as mean $\pm$ standard deviation across houses and base-load realizations: heating-energy use reduction, temperature deviation index, and deviation period index for uncontrolled and DLC houses in the 23-node distribution network.}
\label{tab:comfort_TS2_23node}
\begin{tabularx}{13cm}{@{}
    >{\hsize=0.20\hsize\raggedright\arraybackslash}X
    >{\hsize=0.20\hsize\raggedright\arraybackslash}X
    >{\hsize=0.14\hsize\raggedright\arraybackslash}X
    >{\hsize=0.25\hsize\centering\arraybackslash}X
    >{\hsize=0.12\hsize\centering\arraybackslash}X
    >{\hsize=0.12\hsize\centering\arraybackslash}X
@{}} 
\toprule
HP penetration (\%)  & Control regime & DLC rate (\%) & Energy use reduction (\%) & $T^{\text{dev}}$ (\textdegree C) & $P^{\text{dev}}$ (\%) \\ \midrule
-- & Uncontrolled houses  & --- & --- & $0.02 \pm 0.02$ & $0.0 \pm 0.0$ \\
\midrule
\multirow{2}{*}{25} & ST & 25  & $-18.0 \pm 4.2$ & $0.67 \pm 0.05$ & $70.0 \pm 5.1$ \\
                    & ST & 50  & $-19.6 \pm 4.1$ & $0.67 \pm 0.05$ & $69.9 \pm 5.2$ \\
\midrule
\multirow{8}{*}{50} & ST & 25  & $-11.8 \pm 10.1$ & $0.66 \pm 0.06$ & $66.9 \pm 7.6$ \\
                    & ST+BESS & 25 & $-12.6 \pm 10.6$ & $0.66 \pm 0.06$ & $67.2 \pm 8.2$ \\

                    & ST & 50 & $-16.3 \pm 8.9$ & $0.66 \pm 0.07$ & $66.5 \pm 8.9$ \\

                    & ST+BESS & 50 & $-17.4 \pm 8.1$ & $0.69 \pm 0.07$ & $68.7 \pm 7.8$ \\
                    
                    & ST-DA & 50 & $-18.0 \pm 3.6$ & $0.62 \pm 0.10$ & $63.3 \pm 14.5$ \\

                    & ST$_{\text{ERH-bkp}}$ & 50 & $-21.3 \pm 3.2$ & $0.71 \pm 0.05$ & $70.8 \pm 4.9$ \\

                    & ST$_{\text{VR-local}}$ & 50 & $-16.5 \pm 8.4$ & $0.65 \pm 0.08$ & $66.0 \pm 9.0$ \\
                    
                    & ST$_{\text{VR-DA}}$ & 50 & $-13.5 \pm 9.3$ & $0.62 \pm 0.07$ & $62.8 \pm 9.5$ \\        
\bottomrule
\end{tabularx}
\end{table*}

The ST control raises the indoor temperature deviation well above that of the uncontrolled houses while still reducing the average energy use, as summarized in Table~\ref{tab:comfort_TS2_23node}. Buildings under DLC deviate by about 0.62 to 0.71\textdegree C on average and exceed the 0.5\textdegree C threshold for 62.8 to 70.8\% of the day, while uncontrolled buildings barely deviate from the setpoint ($T^{\text{dev}}=0.02$\textdegree C, $P^{\text{dev}}=0\%$), yet the indoor temperature remains within the admissible $[19,21]$\textdegree C band in all cases.

Table~\ref{tab:results_23node_cop} presents the HP COP, the effective COP, the ERH share and the effective COP of the DLC houses related to the same houses without control ($\Delta \text{COP}$) across control regimes. The DLC of households leaves the effective heating COP nearly unchanged, with $\Delta$COP from $-3.0\%$ to $+6.5\%$ across the control regimes, confirming that its energy-use reduction stems mainly from reduced thermal demand (the temperature setback within the comfort band) rather than from higher conversion efficiency. The HP COP itself is nearly constant (2.06--2.08), so the effective COP differences are driven entirely by the ERH share rather than by shifting heating to higher-COP hours. The effective COP is lowered by resistance-heating use, for both preheating and when the HP capacity is insufficient during the coldest periods, with an ERH share of 3.6\%--20.8\%. The proposed DLC framework therefore delivers demand flexibility at a low heating-efficiency penalty, and even improves efficiency at lower HP penetration and higher DLC rates, with a small penalty only at high penetration with limited DLC rate.
\begin{table*}[htbp!]
\centering
\setlength{\tabcolsep}{2pt}
\scriptsize
\caption{Heating-system efficiency for the 23-node distribution network. Metrics are for the controlled (DLC) houses; the BASE row is the same houses without control. Effective COP is delivered heat divided by total HP+ERH electricity; ERH share is the fraction supplied by the (COP\,$=$\,1) ERH; $\Delta$COP is the effective COP of the controlled houses relative to the same houses without control.}
\label{tab:results_23node_cop}
\begin{tabularx}{14cm}{@{}
    >{\hsize=0.20\hsize\raggedright\arraybackslash}X
    >{\hsize=0.16\hsize\raggedright\arraybackslash}X
    >{\hsize=0.14\hsize\raggedright\arraybackslash}X
    >{\hsize=0.13\hsize\centering\arraybackslash}X
    >{\hsize=0.16\hsize\centering\arraybackslash}X
    >{\hsize=0.15\hsize\centering\arraybackslash}X
    >{\hsize=0.25\hsize\centering\arraybackslash}X
@{}}
\toprule
HP penetration (\%) & Control regime & DLC rate (\%) & HP COP &
 Effective COP & ERH share (\%) &
 $\Delta$COP vs.\ no DLC (\%) \\
\midrule
\multirow{3}{*}{25}  & BASE     &  ---  & 2.06 & 1.91 & 14.0 & --- \\
                     & ST       &  25  & 2.08 & 1.97 &  9.3 & $+3.2$ \\
                     & ST       &  50  & 2.08 & 2.00 &  6.8 & $+4.4$ \\
\midrule
\multirow{9}{*}{50}  & BASE     &  ---  & 2.06 & 1.91 & 14.1 & --- \\
                     & ST       &  25  & 2.08 & 1.86 & 20.8 & $-3.0$ \\
                     & ST+BESS  &  25  & 2.08 & 1.88 & 18.9 & $-1.9$ \\

                     & ST       &  50  & 2.08 & 1.94 & 13.2 & $+1.1$ \\     

                     & ST+BESS  &  50  & 2.08 & 1.96 & 11.2 & $+2.4$ \\
                     & ST-DA & 50 & 2.08 & 1.97 & 10.3 & $+2.7$ \\     

                     & ST$_{\text{ERH-bkp}}$ & 50 & 2.08 & 2.04 & 3.6 & $+6.5$ \\ 

                     & ST$_{\text{VR-local}}$ & 50 & 2.07 & 1.93 & 13.1 & $+1.0$ \\
                     
                     & ST$_{\text{VR-DA}}$     &  50  & 2.08 & 1.88 & 18.0 & $-1.7$ \\                     
\bottomrule
\end{tabularx}
\end{table*}

Dispatching the ERH for preheating alongside the HP, rather than reserving it as a comfort-floor backstop as in ST$_{\text{ERH-bkp}}$, trades heating efficiency for peak reduction. The active use of the ERH lowers the feeder peak to 7.7~MVA against 8.0~MVA for ST$_{\text{ERH-bkp}}$ (see Fig.~\ref{fig:peak_demand_23node}), while raising the ERH share from 3.6\% to 13.2\% and lowering the effective COP from 2.04 to 1.94 (see Table~\ref{tab:results_23node_cop}). The backstop regime instead removes more heating energy, at 21.3\% against 16.3\% (see Table~\ref{tab:comfort_TS2_23node}), by never preheating with resistance heat. The proposed framework therefore prioritizes network relief, calling on low-COP resistance heat only when it buys peak reduction.

\subsubsection{Rolling Horizon Co-Optimization of the Voltage Regulation Devices}

\begin{figure}[pos=htbp!]
    \centering
    \includegraphics[width=1.0\linewidth]{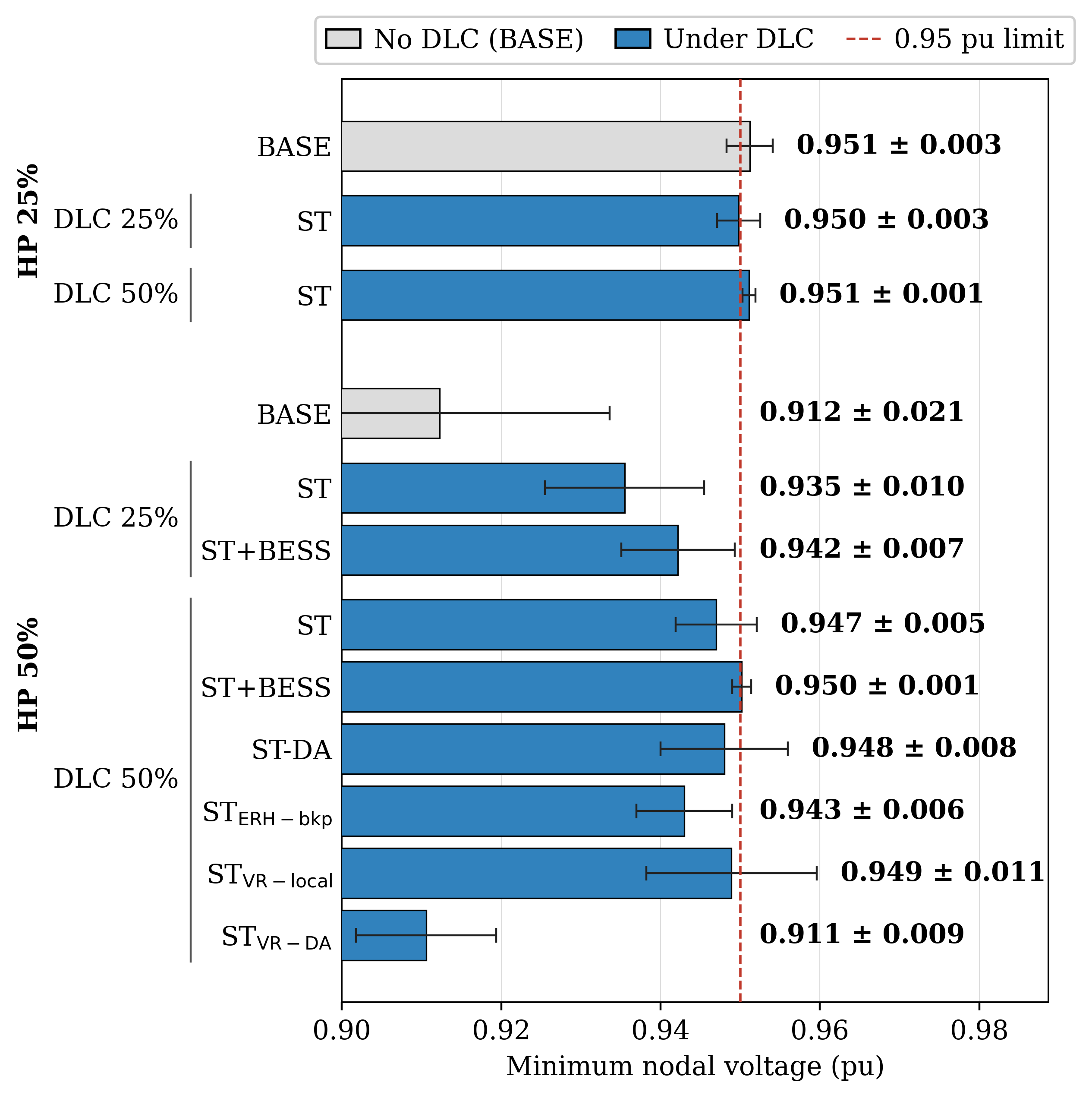}
        \caption{Minimum nodal voltage at the MV level across control regimes for the 23-node distribution network.}
    \label{fig:nodal_voltage_23node}
\end{figure}

Fig.~\ref{fig:nodal_voltage_23node} presents the minimum nodal voltage at the MV level across control regimes. The ST control regime recovers the minimum MV nodal voltage close to the ANSI~C84.1~\cite{ANSI_C84_1_2020} limit of 0.95~pu, whereas day-ahead scheduling of the voltage-regulation devices leaves a voltage violation. At 50\% HP penetration, the proposed ST and ST+BESS regimes recover the minimum nodal voltage to approximately 0.94~pu at 25\% DLC rate, and to approximately 0.95~pu at 50\% DLC rate. The control regime which schedules the voltage regulation day-ahead (ST$_{\text{VR-DA}}$) results in a minimum nodal voltage of 0.91~pu, which is close to the emergency voltage magnitude limit of 0.90~pu from ANSI~C84.1. Throughout, the maximum nodal voltage stays within limits under the hard 1.05~pu constraint, whereas the minimum voltage is not enforced by a hard constraint, to preserve feasibility in the undervoltage cases.

Centrally co-optimizing the OLTC and SCB with the demand-side flexibility, rather than leaving them on their autonomous local controls as in ST$_{\text{VR-local}}$, achieves a similar voltage recovery at a lower cost. Both regimes keep the minimum nodal voltage near the 0.95~pu limit, at 0.947~pu for ST and 0.949~pu for ST$_{\text{VR-local}}$, as shown in Fig.~\ref{fig:nodal_voltage_23node}. However, the centralized coordination lowers the feeder peak to 7.7~MVA against 8.0~MVA, as shown in Fig.~\ref{fig:peak_demand_23node}, and the loss cost to \$825.6 against \$842.6, as reported in Table~\ref{tab:results_TS2_combined_23node}, yielding additional peak and loss reductions that the autonomous local response does not achieve.

\subsubsection{Battery Energy Storage System Operation}

\begin{figure}[pos=htbp!]
    \centering
    \includegraphics[width=1.0\linewidth]{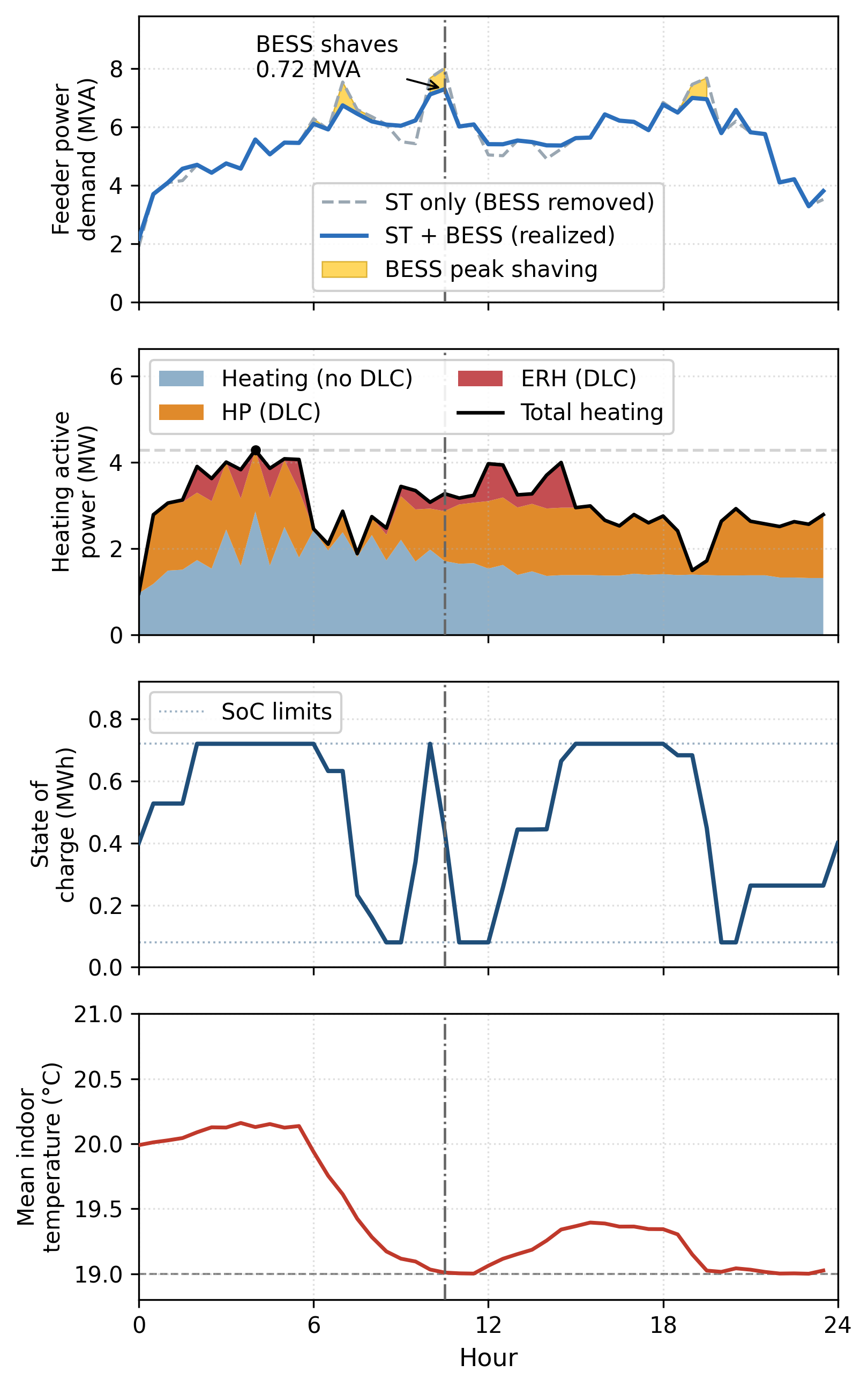}
        \caption{BESS operation coordinated with ST setpoints (ST+BESS) for 50\% HP penetration at 50\% DLC rate in the 23-node distribution network for a single representative day. }
\label{fig:ess_operation_HP50_23node}
\end{figure}

Fig.~\ref{fig:ess_operation_HP50_23node} illustrates the operation of the BESS, its peak-shaving contribution, and its coordination with the heating demand and ST setpoints in the 23-node distribution network for a representative day with 50\% HP penetration and 50\% DLC rate. At each peak the thermostats set back, while the BESS discharges to shave, approximately 0.8~MW at the morning peak (7:00), 0.72~MW at the 10:30 midday peak (the daily maximum, lowering the feeder from 8.0 to 7.3~MVA), and 0.74~MW at the evening peak (19:30), with comfort kept within the [19,21]\textdegree C band throughout. Across the realized days the ST regime provides a daily-peak reduction of 2.31~$\pm$~0.30~MVA (23.1\% of the BASE peak) against a marginal 0.40~$\pm$~0.17~MVA (4.0\%) from the BESS, the thermostats delivering approximately 85\% of the total 2.72~$\pm$~0.44~MVA (27.1\%) reduction. Coordinating the BESS makes the peak-shaving outcome depend less on the DLC rate, while the BESS sizing sets the ceiling of its own peak-shaving contribution within the coordination. 

\subsection{Application Example: 733-Node Distribution Network}

Table~\ref{tab:results_733node_peak_of} presents a summary of the results for the 733-node network at the different penetration levels of HP and DLC rates, along with the maximum loading of the branches and MV/LV transformers. The objective function values are reduced with the ST and ST+BESS regimes, which result in an average maximum branch loading below 100\% for 50\% HP penetration with the ST and ST+BESS regimes at 50\% DLC rate, while the most overloaded transformer remains above 100\% loading across all control regimes, except for the 0\% HP penetration level. The objective function is dominated by the cost of power losses, which accounts for at least 99.9\% of the total objective value across the regimes.
\begin{table*}[htbp!]
\centering
\setlength{\tabcolsep}{2pt}
\scriptsize
\caption{Objective function values and maximum loading of the branch and MV/LV transformers across control regimes for the 733-node distribution network.}
\label{tab:results_733node_peak_of}
\begin{tabularx}{17cm}{@{}
    >{\hsize=0.8\hsize\raggedright\arraybackslash}X
    >{\hsize=0.6\hsize\raggedright\arraybackslash}X
    >{\hsize=0.5\hsize\raggedright\arraybackslash}X
    >{\hsize=0.5\hsize\centering\arraybackslash}X
    >{\hsize=1.1\hsize\centering\arraybackslash}X
    >{\hsize=1.5\hsize\centering\arraybackslash}X
@{}}
\toprule
HP penetration (\%) & Control regime & DLC rate (\%) & OF (\$) & Maximum branch loading (\%) & Maximum MV/LV transformer loading (\%) \\
\midrule
0 & BASE & --- & $38.4 \pm 6.1$ & $44.5 \pm 4.1$ & $82.6 \pm 11.3$ \\
\midrule
\multirow{3}{*}{25}
 & BASE & ---  & $159.8 \pm 10.1$ & $75.4 \pm 5.5$ & $104.9 \pm 8.5$ \\
 & ST   & 25   & $147.8 \pm 8.7$  & $66.8 \pm 4.1$ & $106.1 \pm 8.2$ \\
 & ST   & 50   & $140.3 \pm 7.7$  & $62.8 \pm 2.3$ & $104.1 \pm 10.7$ \\
\midrule
\multirow{5}{*}{50}
 & BASE    & ---  & $413.1 \pm 19.3$ & $113.6 \pm 3.4$ & $152.5 \pm 5.6$ \\
 & ST      & 25   & $374.2 \pm 19.2$ & $104.1 \pm 4.1$ & $130.2 \pm 4.3$ \\
 & ST+BESS & 25   & $372.3 \pm 20.2$ & $99.2 \pm 3.7$ & $130.6 \pm 4.2$ \\

 & ST      & 50   & $339.6 \pm 18.7$ & $98.8 \pm 6.2$  & $130.7 \pm 6.8$ \\
 & ST+BESS & 50   & $338.0 \pm 19.9$ & $93.5 \pm 4.1$ & $130.3 \pm 7.3$ \\
\bottomrule
\end{tabularx}
\end{table*}

\begin{figure}[pos=htbp!]
    \centering
    \includegraphics[width=1.0\linewidth]{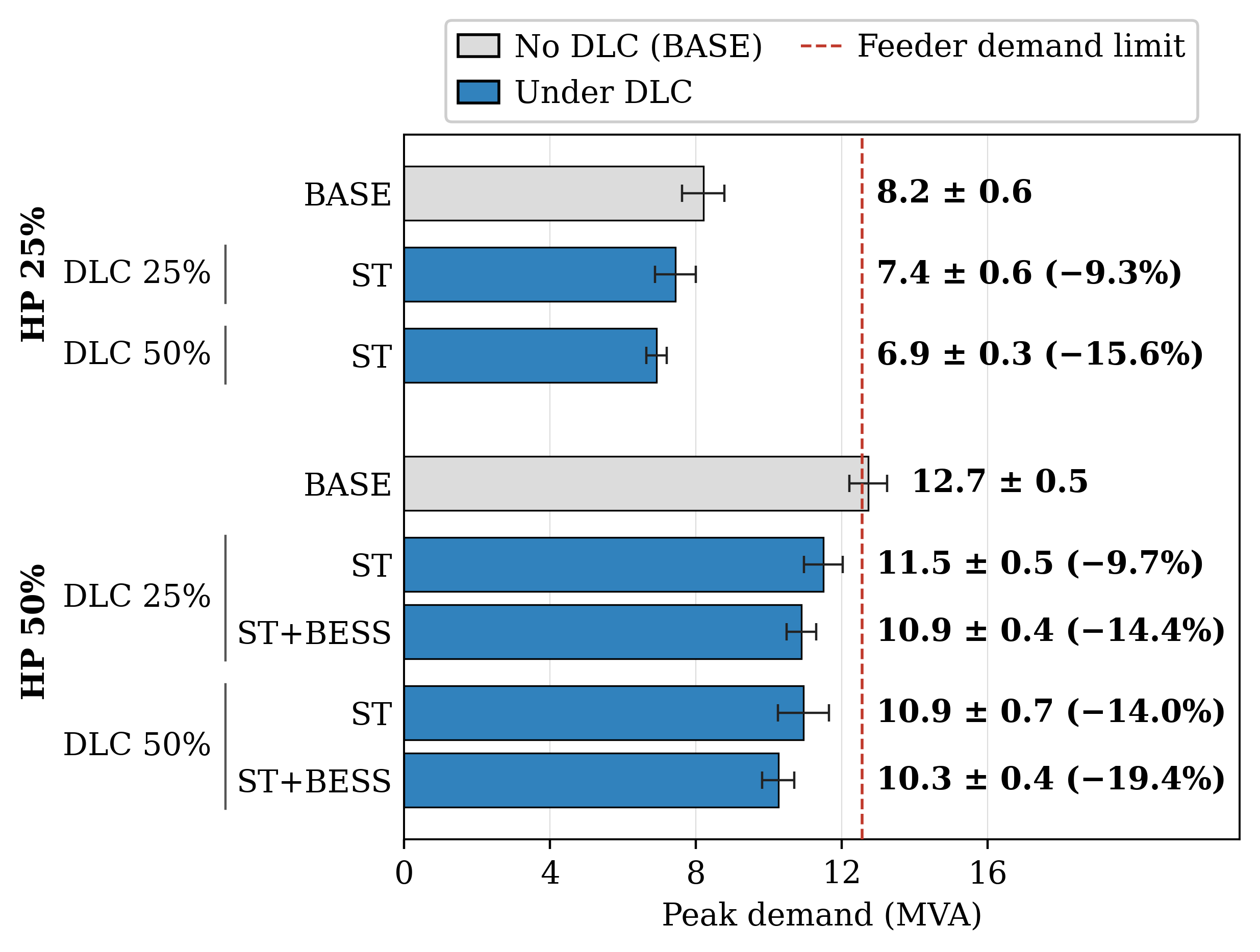}
        \caption{Peak demand across control regimes for the 733-node distribution network.}
    \label{fig:peak_demand_733node}
\end{figure}

Fig.~\ref{fig:peak_demand_733node} presents the peak demand across the control regimes for the 733-node network. At 50\% HP penetration, the ST regime at a 25\% DLC rate reduces the peak demand below the feeder power demand limit of 12.55~MVA, while the ST+BESS regime at a 50\% DLC rate further reduces the peak demand by 19.4\%, to 10.3~MVA.

\begin{figure}[pos=htbp!]
    \centering
    \includegraphics[width=1.0\linewidth]{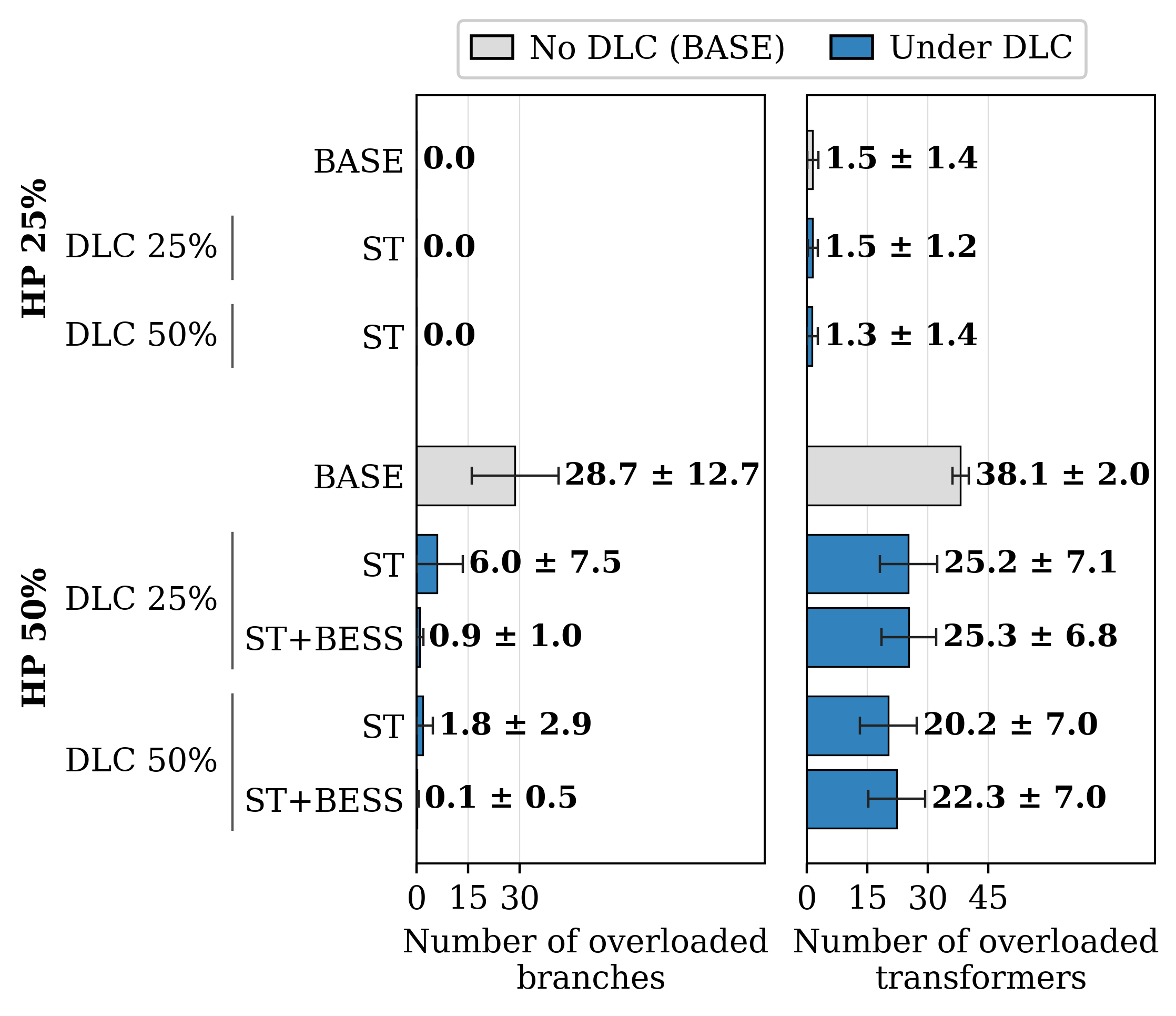}
        \caption{Number of overloaded branches and MV/LV transformers across control regimes for the 733-node distribution network.}
    \label{fig:overloaded_counts_733node}
\end{figure}

Fig.~\ref{fig:overloaded_counts_733node} shows the number of overloaded branches and MV/LV transformers in all control regimes. The ST and ST+BESS regimes significantly reduce the number of overloaded branches and MV/LV transformers under 50\% HP penetration level. An average of 22.7 branches experience overload relief under the ST regime at 25\% DLC rate, increasing to 26.9 branches (93.7\% reduction compared with BASE) at 50\% DLC rate. The number of overloaded MV/LV transformers falls from 38.1 to 20.2 at 50\% DLC rate, a 47\% reduction. The ST+BESS regime nearly eliminates branch overloads, reducing the number of overloaded branches to 0.9 and 0.1 at 25\% and 50\% DLC rates, respectively.

\subsection{Computational Performance of the Proposed MILP-based MPC Framework}

The maximum computational time required to obtain the optimal solution at each MPC time step, considering all base-load realizations for each HP penetration level and DLC rate, is reported in Table~\ref{tab:comp_time}. Simulations were performed on an Intel Core i7 2.10~GHz processor with 32~GB of RAM using 12~threads. The maximum solution time generally increases with HP penetration and the DLC rate, as well as when the BESS is included in the co-optimization, for both distribution networks. 

\begin{table}[!t]
\centering
\setlength{\tabcolsep}{2pt}
\scriptsize
\caption{Maximum computational time (s) across HP penetration levels and DLC rates, considering the BASE, ST and ST+BESS control regimes, where values in parentheses correspond to ST+BESS.}
\label{tab:comp_time}
\begin{tabularx}{8cm}{@{}
    >{\hsize=0.15\hsize\raggedright\arraybackslash}X
    >{\hsize=0.30\hsize\raggedright\arraybackslash}X
    >{\hsize=0.10\hsize\centering\arraybackslash}X
    >{\hsize=0.20\hsize\centering\arraybackslash}X 
    >{\hsize=0.20\hsize\centering\arraybackslash}X
    @{}} 
\toprule
\multirow{2}{*}{Network} & 
\multirow{2}{*}{HP penetration (\%)} & \multirow{2}{*}{BASE} & 
\multicolumn{2}{c}{DLC rate (\%)} \\
\cmidrule(lr){4-5}
& & & 25 & 50 \\
\midrule
\multirow{3}{*}{23-node}
& 0  & 0.7 & --- & --- \\
& 25 & 0.7 & 10.4 & 45.6 \\
& 50 & 0.9 & 42.5 (151.0) & 174.8 (697.2) \\
\midrule
\multirow{3}{*}{733-node}
& 0  & 4.8 & --- & --- \\
& 25 & 5.3 & 14.1 & 32.7 \\
& 50 & 3.5 & 87.3 (113.3) & 248.7 (762.1) \\
\bottomrule
\end{tabularx}
\end{table}

\section{Conclusions}
\label{sec:conclusions}

Coordinating households' heating flexibility with the existing voltage-regulation assets and BESS allows MV distribution feeders to host higher HP electrification without reinforcement. To this end, this paper proposes a MILP-based MPC that co-optimizes, over a rolling horizon, the ST setpoints of households heating with HPs and ERHs, the voltage-regulation devices (OLTC, SVR, and SCB), and grid-scale BESS. Evaluated on a 23-node and a real 733-node network across HP penetration levels, the coordinated DLC of ST lowered the feeder peak demand, relieved branch and transformer congestion, and recovered nodal voltages, while keeping indoor temperatures within the comfort band and reducing end-user heating energy use. The binding constraint differed by feeder: undervoltage on the long, high-impedance 23-node network, where ST control recovered the minimum voltage to the statutory limit, and thermal congestion on the 733-node network, where it cleared nearly all branch and transformer overloads.  

The comparison against restricted control regimes suggests practical outcomes for DR programs of ST. Re-optimizing the ST setpoints over the rolling horizon outperformed a day-ahead ST schedule, as the controller re-dispatches the preheating and setback signals to the realized demand. Dispatching the ERH for preheating, rather than reserving it as a comfort backup, deepened the peak reduction at a small heating-efficiency cost. Co-optimizing the OLTC and SCB with the demand-side flexibility achieved a voltage recovery similar to their autonomous local control while yielding additional peak and loss reductions, whereas committing these devices day-ahead left the minimum voltage near its emergency limit under the realized loading. The BESS further reduced the peak demand and branch loading by charging and discharging in coordination with the preheating and setback signals. For distribution utilities, these results indicate that existing ST DR programs can evolve from peak-event curtailment into a network-operation resource when the setpoints are dispatched together with the voltage-regulation devices and storage that the utility operates.

Future work includes an unbalanced network representation to assess the voltage unbalance at the supply points, and a robust or chance-constrained formulation of the voltage-regulation control to strengthen undervoltage mitigation under forecast uncertainty, at the cost of additional computation for rolling-horizon operation.

\appendix

\section{Autonomous Operation of OLTC and SCB}
\label{app:autonomous_OLTC_SCB}

The OLTC operates autonomously through LDC~\cite{Kersting20171} using active part of the current leaving the substation $(f_{1-2,t}^{\text{Re}}$), mapping it in \eqref{eq:olt_ldc_target} to a continuous voltage target ($V_{t}^{\text{tgt}}$) that a half-step deadband quantizes to the nearest physical tap with saturation slacks $s_{t}^{\text{hi}},s_{t}^{\text{lo}} \ge 0$, where $R_{\text{LDC}}$ is the compensator resistance setting and $V_{\text{LDC}}^{\text{set}}$ is the base setpoint. Constraints \eqref{eq:oltc_ldc_band_hi} and \eqref{eq:oltc_ldc_band_lo} force $V_{t}^{\text{ss}}$ onto the tap closest to $V_{t}^{\text{tgt}}$, while \eqref{eq:oltc_ldc_sat_hi} and \eqref{eq:oltc_ldc_sat_lo} allow the tap to rest at the bottom or top when the target leaves the regulation range.   
\begin{subequations}
\label{eq:oltc_ldc}
\begin{align}
    & V_{t}^{\text{tgt}} = V_{\text{LDC}}^{\text{set}} + R_{\text{LDC}} f_{1-2,t}^{\text{Re}} 
    \label{eq:olt_ldc_target} \\
    &  V_{t}^{\text{ss}} - V_{t}^{\text{tgt}} \leq \frac{\Delta V ^{\text{OLTC}}}{2} + b_{t}^{\text{hi}} \label{eq:oltc_ldc_band_hi} \\ 
    &  V_{t}^{\text{tgt}} - V_{t}^{\text{ss}}  \leq \frac{\Delta V ^{\text{OLTC}}}{2} + b_{t}^{\text{lo}}
    \label{eq:oltc_ldc_band_lo} \\
    &  s_{t}^{\text{hi}} \leq M \delta_{0,t} \label{eq:oltc_ldc_sat_hi} \\ 
    &  s_{t}^{\text{lo}} \leq M \delta_{N_{\text{taps}}^{\text{OLTC}},t} \label{eq:oltc_ldc_sat_lo}
\end{align}
\end{subequations}

The SCB is switched autonomously based on the voltage magnitude at its own node, emulating a local voltage-controlled switching rule. Its operation retains the level selection and constant-impedance injection of~\eqref{eq:scb_model}~\cite{Resener2016895}, to which the local switching logic in~\eqref{eq:scb_local} is added. A binary saturation indicator $o_{m,t}^{\text{sat}}$ can equal $1$ only when all SCB levels are switched in, as imposed by~\eqref{eq:scb_sat}. Constraint~\eqref{eq:scb_volt} then requires the nodal voltage $V_{m,t}$ to remain at or above the switching threshold $V_{\text{low}}^{\text{scb}}$, a requirement relaxed through the big-M term only when the bank is saturated and no further level can be added. A small penalty $\lambda^{\text{scb}}\sum_{l,t} l\,c_{m,l,t}$ is added to the objective function so that the bank switches in only the fewest levels needed to meet the threshold. This penalty is excluded from the reported objective values.
\begin{subequations}
\label{eq:scb_local}
\begin{align}
    & \sum_{l\in C_{m}^{\text{scb}}} l\, c_{m,l,t} \geq C_{m}^{\text{scb}}\, o_{m,t}^{\text{sat}} \label{eq:scb_sat} \\
    & V_{m,t} \geq V_{\text{low}}^{\text{scb}} - M\, o_{m,t}^{\text{sat}} \label{eq:scb_volt}
\end{align}
\end{subequations}

The OLTC droop is calibrated for the range of $V^{\text{ss,min}}=1.00$ and $V^{\text{ss,max}}=1.05$, for an operating band between a light-load of $I_{\text{min}}=54$~A and a heavy-load $I_{\text{max}}=301$~A design peak. This gives a slope of $R_{\text{LDC}}=0.068$~pu/pu and base point of $V_{\text{LDC}}^{\text{set}}=0.989$~pu, with a tap deadband of $0.00625$~pu.   

For the 23-node distribution network, the SCB at node~$21$ holds its bus voltage magnitude at or above $V_{\text{low}}^{\text{scb}}=0.98$~pu, with a switching penalty $\lambda^{\text{scb}}=1.0$ per level per step.

\section{Control of ST with ERH as Comfort-Floor Backstop Only (ST$_{\text{ERH-bkp}}$)}
\label{app:ST_ERH_bkp}

In \textit{ST$_{\text{ERH-bkp}}$}, activating any ERH stage ($u_{e,s,t}^{\text{erh}}=1$) pins the commanded setpoint to the lower comfort bound~\eqref{eq:erhbk_model}. Since $T_{e,t}^{\text{ind,set}}\ge T_{\min}^{\text{ind}}$, this forces $T_{e,t}^{\text{ind,set}}=T_{\min}^{\text{ind}}$ whenever any stage is on.
\begin{equation}
    T_{e,t}^{\text{ind,set}} \le T_{\min}^{\text{ind}}
    + \big(T_{\max}^{\text{ind}} - T_{\min}^{\text{ind}}\big)\big(1 - u_{e,s,t}^{\text{erh}}\big)  
    \label{eq:erhbk_model}
\end{equation}

\printcredits


%
\section*{Declaration of generative AI and AI-assisted technologies in the manuscript preparation process }

During the preparation of this work the author(s) used Claude and ChatGPT in order to assist in refining the clarity and grammar. All content generated by the tool was reviewed and revised by the authors. No AI tools were used in the analysis, interpretation of data, or generation of original research findings. The authors are solely responsible for the research design, data interpretation, and conclusions drawn.

\bibliographystyle{unsrtnat}

\bibliography{cas-refs}



\end{document}